# A Generalized Probabilistic Monitoring Model with Both Random and Sequential Data

Wanke Yu[a], Min Wu[a], Biao Huang[b], and Chengda Lu[a]

[a] The School of Automation, China University of Geosciences, Wuhan 430074, China, with Hubei Key Laboratory of Advanced Control and Intelligent Automation for Complex Systems, Wuhan 430074, China, and also with Engineering Research Center of Intelligent Technology for Geo-Exploration, Ministry of Education, Wuhan 430074, China

[b] The Department of Chemical and Materials Engineering, University of Alberta, Edmonton, AB T6G 2G6, Canada

**Abstract**

Many multivariate statistical analysis methods and their corresponding probabilistic counterparts have been adopted to develop process monitoring models in recent decades. However, the insightful connections between them have rarely been studied. In this study, a generalized probabilistic monitoring model (GPMM) is developed with both random and sequential data. Since GPMM can be reduced to various probabilistic linear models under specific restrictions, it is adopted to analyze the connections between different monitoring methods. Using expectation maximization (EM) algorithm, the parameters of GPMM are estimated for both random and sequential cases. Based on the obtained model parameters, statistics are designed for monitoring different aspects of the process system. Besides, the distributions of these statistics are rigorously derived and proved, so that the control limits can be calculated accordingly. After that, contribution analysis methods are presented for identifying faulty variables once the process anomalies are detected. Finally, the equivalence between monitoring models based on classical multivariate methods and their corresponding probabilistic graphic models is further investigated. The conclusions of this study are verified using a numerical example and the Tennessee Eastman (TE) process. Experimental results illustrate that the proposed monitoring statistics are subject to their corresponding distributions, and they are equivalent to statistics in classical deterministic models under specific restrictions.

## 1. Introduction

In the past decades, process monitoring with multivariable analysis methods has been intensively studied in process control community and extensively applied in industrial plants (Kruger, Kumar, Littler, 2007; Qin, 2012; Shang et al., 2017). By characterizing the distribution of archived historical data collected in nominal operation conditions, monitoring models can be effectively developed using theory of testing hypothesis (Zhao et al., 2021; Yu, Zhao, 2019). Based on the obtained models, alarms can then be timely raised for process anomalies that transgress the routine data distribution (Alcala, Qin, 2009; Chen, Zhao, 2021; Yu, Zhao, 2020; Zhao, Huang, Liu, 2016).

Based on the collinearity, multivariate statistical analysis methods are adopted to develop process monitoring models by projecting process data into low-dimensional subspaces. The commonly used methods include principal component analysis (PCA) (Li et al., 2000), canonical correlation analysis (CCA) (Russell et al., 2000), slow feature analysis (SFA) (Wiskott, Sejnowski, 2002; Shang et al., 2015), etc. Among them, PCA is developed by decomposing the process data into a principal subspace with maximized variability and a residual subspace with reconstruction error (Li et al., 2000). CCA can maximize the correlation between latent variables derived from input and output samples, and thus the developed model consists of information of two datasets (Russell et al., 2000). SFA extracts slowly varying latent variables from sequential data, and develops a monitoring model with both static and temporal information. In this way, it can effectively distinguish operating condition deviations and process dynamic anomalies (Shang et al., 2015). Once a process anomaly is detected, the faulty variables should be isolated to identify the section where fault occurs. In general, contribution analysis methods are the most commonly used methods, which are carried out by evaluating the contribution of each variable to the monitoring statistics (Alcala, Qin, 2011; Shang, et al., 2016). As summarized by Alcala & Qin (2011), contribution analysis methods can be roughly divided into three categories, namely, the general decomposition contribution (GDC), the reconstruction based contribution (RBC), and the diagonal contribution (DC).

Although the aforementioned multivariate statistical analysis methods are easy to be implemented, they are deterministic models and have not taken the additive noise into consideration (Raveendran, Huang, 2020; Scott et al., 2020). That is, only the estimations of the model parameters can be obtained, but the confidence levels of obtained models are ignored. To solve this issue, many probabilistic graphic models (PGMs) are proposed, including probabilistic PCA (PPCA) (Tipping et al., 1999a; Tipping et al., 1999b), probabilistic CCA (PCCA) (Bach, Jordan, 2005), probabilistic SFA (PSFA) (Shang et al., 2015; Zafeiriou et al., 2016), etc. Instead of singular value decomposition (SVD), PGMs are generally established using expectation maximization (EM) algorithm (Moon, 1996). As a result, a better interpretation to the latent variable model is depicted and the missing values in process data can also be effectively complemented (Kim & Lee, 2003). Similar to multivariate methods, many PGMs are also introduced to develop monitoring models. Chen et al. (2009) proposed a

This work was supported in part by the National Natural Science Foundation of China under Grants 61733016 and 62103387; in part by the National Key R&D Program of China under Grant 2018YFC0603405; and in part by the 111 Project under Grant B17040. Corresponding author: Min Wu.

Email Addresses: yuwanke@cug.edu.cn (Wanke Yu), wumin@cug.edu.cn (Min Wu), biao.huang@ualberta.ca (Biao Huang), luchengda@cug.edu.cn (Chengda Lu).

robust PPCA based monitoring strategy with missing data, which is robust to the presence of outliers by replacing Gaussian distribution with heavy-tailed *t*-distribution. Guo et al. (2016) designed a monitoring model using PSFA, which is developed for sequential data based on a Markov chain.

In recent years, the connection between classical methods and their probabilistic counterparts has received attentions. Based on a linear system, Raveendran et al. (2018) proposed a generalized probabilistic linear latent variable model (GPLLVM). Specifically, GPLLVM can be reduced to PPCA (Tipping, Bishop, 1999a) and PCCA (Bach, Jordan, 2005) by setting restrictions on the model parameters. Besides, the equivalence between statistics of GPLLVM with specific restrictions and classical deterministic methods (i.e., PCA and CCA) has also been proved strictly and in detail. However, GPLLVM is designed to model sequentially uncorrelated data and the latent variables derived from input and output samples are assumed to be identical. As we know, the operation status of current process is generally affected by its previous samples (Dong, Qin, 2018; Zhao, Huang, 2018). That is, the temporal information within the process data should also be taken into consideration. Furthermore, the flexibility of the GPLLVM may be restricted by the assumption that the latent variables of input and output data are identical.

In this study, a generalized probabilistic monitoring model (GPMM) is proposed for developing models with both random and sequential data. Here, random data indicates sequentially uncorrelated data or process data whose sequential information has not been taken into consideration. In comparison with GPLLVM, the proposed method has a greater flexibility. Specifically, it subsumes more probabilistic linear latent variable models, including PPCA, PCCA and PSFA. For both random and sequential cases, the model parameters of GPMM are estimated and provided using EM algorithm. As a comparison, the M-step in both cases are almost the same, but their E-step are totally different since random case is developed using current samples and sequential case is established based on entire process data. Based on the obtained model, a series of statistics are designed to monitor different aspects of the process system. The distributions of these statistics are strictly derived and proved, and thus the corresponding control limits can be calculated accordingly. After that, contribution analysis methods have been analyzed for identifying the faulty variables. Finally, the equivalence between monitoring statistics of GPMM with specific restrictions and classical multivariate statistical methods is demonstrated. The contributions of this study are summarized as below

1) A generalized process monitoring model is proposed with both random and sequential data, and the estimations of model parameters for both cases are provided using EM algorithm.

2) Based on the obtained model, statistics are designed for monitoring different aspects of the process system, and their distributions are derived to calculate the control limits.

3) The equivalence between monitoring statistics derived from GPMM and conventional multivariate methods (i.e., PCA, CCA and SFA) are rigorously proved.

## 2. Preliminaries

### 2.1 PCA based Monitoring Strategy

PCA is a dimensionality reduction method which is used to capture the variance of the data. Define the output of a system as $\mathbf{Y} = \{y_1, y_2, \cdots, y_T\} \in \Re^{p \times T}$, which has been normalized to have zero mean and unit variance. Based on SVD method, its covariance matrix $\mathbf{\Psi} = \mathbf{Y}\mathbf{Y}^T / T$ can be decomposed as below

$$\mathbf{\Psi} = \mathbf{B}\mathbf{\Pi}\mathbf{B}^T \quad (1)$$

where, $\mathbf{\Pi} \in \Re^{p \times p}$ is a diagonal matrix that consists of $p$ eigenvalues. $\mathbf{B}$ is composed of orthonormal eigenvectors as columns corresponding to the eigenvalues in $\mathbf{\Pi}$.

The $T^2$ statistic in PCA method (Li et al., 2000) is the normalized sum of squares of latent variables, Define the eigenvectors with the largest $r$ eigenvalues as $\mathbf{B}_R \in \Re^{p \times r}$. For a normalized new sample $y_{new} \in \Re^{p \times 1}$, its $T^2$ statistic can be calculated as below

$$T^2 = y_{new}^T \mathbf{B}_R \left(\mathbf{\Pi}_R\right)^{-1} \mathbf{B}_R^T y_{new} \quad (2)$$

where, $\mathbf{\Pi}_R$ is a diagonal matrix with the largest $r$ eigenvalues.

The *SPE* statistic is the sum of squares of the reconstruction residuals, and it is given by

$$SPE = y_{new}^T (\mathbf{I}_p - \mathbf{B}_R \mathbf{B}_R^T) y_{new} \quad (3)$$

where, $\mathbf{I}_p \in \Re^{p \times p}$ is an identity matrix.

### 2.2 CCA based Monitoring Strategy

CCA method is developed to maximize the correlation between two sets of variables. Let $\mathbf{X} = \{x_1, x_2, \cdots, x_T\} \in \Re^{q \times T}$ to be the inputs of a system whose outputs are $\mathbf{Y}$. Similar to PCA, the variables in $\mathbf{X}$ and $\mathbf{Y}$ have also been normalized to have zero mean and unit variance.

Define $\mathbf{\Sigma}_{xx}$ and $\mathbf{\Sigma}_{yy}$ as the covariance matrices of $\mathbf{X}$ and $\mathbf{Y}$, respectively. The cross covariance matrix between $\mathbf{X}$ and $\mathbf{Y}$ is defined as $\mathbf{\Sigma}_{xy}$. A matrix decomposition can then be depicted as below using SVD method.

$$\mathbf{\Sigma}_{xx}^{-1/2} \mathbf{\Sigma}_{xy} \mathbf{\Sigma}_{yy}^{-1/2} = \tilde{\mathbf{B}}_x \mathbf{\Pi} \tilde{\mathbf{B}}_y^T \quad (4)$$

where, $\mathbf{\Pi} \in \Re^{r \times r}$ is a diagonal matrix with the singular values, $\tilde{\mathbf{B}}_x \in \Re^{q \times r}$ and $\tilde{\mathbf{B}}_y \in \Re^{p \times r}$ are composed of orthogonal eigenvectors.

Based on Eq. (4), the projection matrices $\mathbf{B}_x$ and $\mathbf{B}_y$ can be calculated as below

$$\mathbf{B}_x = \mathbf{\Sigma}_{xx}^{-1/2} \tilde{\mathbf{B}}_x, \quad \mathbf{B}_y = \mathbf{\Sigma}_{yy}^{-1/2} \tilde{\mathbf{B}}_y \quad (5)$$

For a new input $x_{new} \in \Re^{q \times 1}$ and its corresponding output $y_{new} \in \Re^{p \times 1}$, the monitoring statistics of CCA (Russell et al., 2000) are given by

$$T_x^2 = x_{new}^T \mathbf{B}_x \mathbf{B}_x^T x_{new}, \quad T_y^2 = y_{new}^T \mathbf{B}_y \mathbf{B}_y^T y_{new} \quad (6)$$

### 2.3 SFA based Monitoring Strategy

SFA method is proposed to extract slowly varying latent variables from sequence data (Wiskott et al., 2002). Given the continuous outputs of a system $\mathbf{Y} = \{y_1, y_2, \cdots, y_T\} \in \Re^{p \times T}$, the objective of SFA is to minimize the temporal difference. Define

the covariance matrix as $\Psi = YY^T/T$, and it can be decomposed using SVD as below
$$\Psi = A\Pi A^T \quad (7)$$
Based on Eq. (7), the whitening transformation can be implemented by
$$z_t = \Pi^{-1/2} A^T y_t \quad (8)$$
Define $\Sigma$ as the covariance matrix of the first order difference $\dot{z}_t = z_t - z_{t-1}$. Then, the orthogonal matrix $P$ can be calculated using following equation
$$\Sigma = P\Omega P^T \quad (9)$$
After that, the projection matrix of SFA is given by
$$B = P^T \Pi^{-1/2} A^T \quad (10)$$
Select the projections with $r$ smallest eigenvalues (i.e., slow features) as $B_S \in \Re^{p \times r}$, and the remaining projections (i.e., fast features) are defined as $B_F \in \Re^{p \times (p-r)}$. Then, the static indices of the newly collected sample (Shang et al., 2015) are given by
$$T_S^2 = y_{new}^T B_S^T B_S y_{new}, \quad T_F^2 = y_{new}^T B_F^T B_F y_{new} \quad (11)$$
Besides, the temporal statistics can be calculated using the difference of the obtained slow feature (Shang et al., 2015)
$$S_S^2 = \dot{y}_{new}^T B_S^T B_S \dot{y}_{new}, \quad S_F^2 = \dot{y}_{new}^T B_F^T \Omega_F^{-1} B_F \dot{y}_{new} \quad (12)$$
where, $\Omega_F$ is a diagonal matrix that consists of the $p - r$ largest eigenvalue in $\Omega$.

*2.4 GPLLVM model*

As mentioned in introduction, the GPLLVM method can be reduced to various probabilistic linear models under specific restrictions (Raveendran et al., 2018). Specifically, it is developed to model the outputs $Y \in \Re^{p \times T}$ given the inputs $X \in \Re^{q \times T}$ which are corrupted by noise and the deterministic inputs $C \in \Re^{o \times T}$. The GPLLVM model can be represented as below
$$\begin{aligned} y_t &= Us_t + Fc_t + e_y, \quad e_y \sim N(0, \Lambda_y) \\ x_t &= Vs_t + e_x, \quad e_x \sim N(0, \Lambda_x) \end{aligned} \quad (13)$$
where, $s_t \sim N(0, I_r)$ is the latent variable, and $c_t \in \Re^{o \times 1}$ indicates the deterministic input. $e_y$ and $e_x$ are the noise terms that are independent, identical and normally distributed with covariance matrices $\Lambda_y$ and $\Lambda_x$, respectively. Besides, $U \in \Re^{p \times r}$, $F \in \Re^{p \times o}$, and $V \in \Re^{q \times r}$ are the coefficient matrices of the GPLLVM model.

When there are no inputs and the output noise covariance $\Lambda_y$ is diagonal, the GPLLVM model reduces to probabilistic factor analyzer as below
$$y_t = Uz_t + e_y, \quad e_y \sim N(0, \Lambda_y) \quad (14)$$
When $\Lambda_y$ is isotropic (i.e., $\Lambda_y = \pi^2 I$), the model defined in Eq. (14) can be further reduced to PPCA (Tipping, et al., 1999a) as below
$$y_t = Uz_t + e_y, \quad e_y \sim N(0, \pi^2 I) \quad (15)$$
When there are no deterministic inputs $C$, the GPLLVM model reduces to the PCCA (Bach et al., 2005) as below
$$\begin{aligned} y_t &= Us_t + e_y, \quad e_y \sim N(0, \Lambda_y) \\ x_t &= Vs_t + e_x, \quad e_x \sim N(0, \Lambda_x) \end{aligned} \quad (16)$$

## 3 GPMM Method

*3.1 Formulation of the GPMM*

Consider a process system with inputs $X \in \Re^{q \times T}$ and outputs $Y \in \Re^{p \times T}$. Assume that the latent variables of $y_t$ and $x_t$ are $z_t$ and $s_t$ respectively, and they are closely correlated with each other. A probabilistic linear latent model that models $Y$ given the inputs $X$ can be formulated as
$$\begin{aligned} y_t &= Uz_t + c_y + e_y, \quad e_y \sim N(0, \Lambda_y) \\ x_t &= Vs_t + c_x + e_x, \quad e_x \sim N(0, \Lambda_x) \\ z_t &= Ws_t + \varepsilon, \quad \varepsilon \sim N(0, \Lambda_\varepsilon) \end{aligned} \quad (17)$$
where, $s_t \in \Re^{r \times 1}$ and $z_t \in \Re^{r \times 1}$ are the latent variables derived from input $x_t$ and output $y_t$, respectively. $c_x \in \Re^{q \times 1}$ and $c_y \in \Re^{p \times 1}$ are the deterministic values of the input and output models, respectively. It is noted that the dimensions of latent variables satisfy $r \leq \min\{p, q\}$. $U \in \Re^{p \times r}$ and $V \in \Re^{q \times r}$ are the coefficient matrices, and $W \in \Re^{r \times r}$ is the transition matrix. $e_y$, $e_x$, and $\varepsilon$ are noise terms that are independent, identical and normally distributed with $\Lambda_y$, $\Lambda_x$, and $\Lambda_\varepsilon$, respectively.

Similar to many classical PGMs, latent variable $s_t$ follows $s_t \sim N(0, I_r)$. The transition matrix $W$ and covariance matrix $\Lambda_\varepsilon$ are diagonal, and their relationship is interpreted using Eq. (18). According to the third equation in Eq. (17), latent variables $z_t$ also satisfies $z_t \sim N(0, I_r)$.
$$\begin{aligned} W &= diag\{\lambda_1, \lambda_2, \cdots \lambda_r\} \\ \Lambda_\varepsilon &= diag\{1 - \lambda_1^2, 1 - \lambda_2^2, \cdots 1 - \lambda_r^2\} \end{aligned} \quad (18)$$
where, $\lambda_i \in [0, 1]$ is a parameter which is used to determine the similarity between the $i$-th variable in $s_t$ and $z_t$.

In recent years, an input-output probabilistic SFA (I/O PSFA) is proposed with both input and output information (Fan et al., 2019). It is noted that the formulations of GPMM and I/O PSFA are different, which finally results in different estimations of model parameters. Besides, the objectives of GPMM and I/O PSFA are also different. Specifically, GPMM is proposed to establish an insightful connection between different monitoring models, but I/O PSFA is used to develop inferential models with better predictive ability.

In GPMM model, the input and output latent variables are assumed to be closely correlated, and thus the optimization problem in Eq. (19) can be derived. Based on this equation, it is easy to find that a larger $\lambda_i$ generally implies a stronger correlation between $s_t$ and $z_t$.
$$\begin{aligned} \max_W E_{X,Y,\Theta}\left(s_t^T z_t\right) &= \max_W E_{X,Y,\Theta}\left(W \times s_t^T s_t + s_t^T \varepsilon\right) \\ &= \max_W W \times E_{X,Y,\Theta}\left(s_t^T s_t\right) = \max_W trace(W) \\ s.t. \quad E_{X,Y,\Theta}\left(s_t^T s_t\right) &= 1, \quad E_{X,Y,\Theta}\left(z_t^T z_t\right) = 1 \end{aligned} \quad (19)$$
where, $\Theta$ indicates the model parameters of GPMM.

Similar to GPLLVM method, GPMM method subsumes the probabilistic counterparts of many latent variable methods. Specifically, GPMM can be reduced to GPLLVM when setting

$\mathbf{W} = \mathbf{I}_r$, $\mathbf{c}_x = \mathbf{0}$, and $\mathbf{c}_y = \mathbf{F}\mathbf{c}_t$. Besides, it can also be used to establish models for sequential data.

Although both methods can be reduced to PCCA, there are difference between GPMM and GPLLVM. As we known, CCA method is a constrained optimization problem, which results in the maximization of correlation. This is difficult to be formulated as a transformation relationship between $s_t$ and $z_t$. As an alternative, the similarity between input and output latent variables is adopted in GPMM. In comparison with classical methods (Bach, et al., 2005; Raveendran et al., 2018), which assumes that the input and output latent variables are identical, GPMM has less restrictions in the latent variables and thus it has a greater flexibility.

When the inputs and outputs are the samples of a correlated sequence, GPMM method can also be used to develop models by setting $y_t = x_{t+\tau}$. Correspondingly, the derived latent variables satisfy $z_t = s_{t+\tau}$, and we also have $\mathbf{U} = \mathbf{V}$, $\mathbf{c}_y = \mathbf{c}_x$, and $\mathbf{\Lambda}_y = \mathbf{\Lambda}_x$. For this case, GPMM reduces to multiple Markov chains which share the same parameters. Specifically, the PGM depicted below reduces to PSFA (Shang et al., 2015) when interval satisfies $\tau = 1$ and $\mathbf{\Lambda}_x$ is a diagonal matrix.

$$\begin{aligned}\mathbf{x}_t &= \mathbf{V}\mathbf{s}_t + \mathbf{c}_x + \mathbf{e}_x, \quad \mathbf{e}_x \sim N(0, \mathbf{\Lambda}_x)\\ \mathbf{s}_{t+\tau} &= \mathbf{W}\mathbf{s}_t + \mathbf{\varepsilon}, \quad \mathbf{\varepsilon} \sim N(0, \mathbf{\Lambda}_\varepsilon)\end{aligned} \quad (20)$$

Since two different latent variables are included, the proposed GPMM method can be developed with a greater flexibility. As a result, more process monitoring methods can be analyzed in the same framework instead of being viewed in isolation. Besides, the difference between monitoring models based on random and sequential data can also be further studied in this framework. Furthermore, the proposed GPMM method provides a potential path to develop more monitoring methods. For example, the monitoring model based on the dissimilarity between the input and output latent variables.

*3.2 Solution of GPMM with Random Data*

For illustration, the graphical structure of GPMM with random data is depicted in Figure 1. As shown in this figure, the input $x_t$ and output $y_t$ are conditioned on the latent variables $s_t$ and $z_t$, respectively. Besides, latent variable $z_t$ is conditioned on latent variable $s_t$. For this case, the model parameters are defined as $\mathbf{\Theta} = \{\mathbf{U}, \mathbf{V}, \mathbf{W}, \mathbf{\Lambda}_y, \mathbf{\Lambda}_x, \mathbf{\Lambda}_\varepsilon\}$.

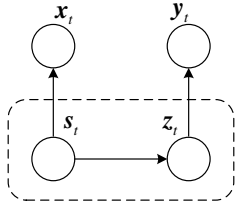

Fig. 1. The graphical structure of GPMM with random data.

The complete data likelihood of GPMM with random data can be shown as below

$$P(\mathbf{X}, \mathbf{Y}, \mathbf{Z}, \mathbf{S} \mid \mathbf{\Theta}) = \prod_{t=1}^{T} P(\mathbf{y}_t \mid \mathbf{z}_t) \times \prod_{t=1}^{T} P(\mathbf{x}_t \mid \mathbf{s}_t) \\ \times \prod_{t=1}^{T} P(\mathbf{z}_t \mid \mathbf{s}_t) \times \prod_{t=1}^{T} P(\mathbf{s}_t) \quad (21)$$

where

$$\begin{aligned}P(\mathbf{y}_t \mid \mathbf{z}_t) &= N(\mathbf{y}_t \mid \mathbf{U}\mathbf{z}_t + \mathbf{c}_y, \mathbf{\Lambda}_y)\\ P(\mathbf{x}_t \mid \mathbf{s}_t) &= N(\mathbf{x}_t \mid \mathbf{V}\mathbf{s}_t + \mathbf{c}_x, \mathbf{\Lambda}_x)\\ P(\mathbf{z}_t \mid \mathbf{s}_t) &= N(\mathbf{z}_t \mid \mathbf{W}\mathbf{s}_t, \mathbf{\Lambda}_\varepsilon)\end{aligned} \quad (22)$$

The Q-function can be formally derived with the conditional expectation of $\ln P(\mathbf{X}, \mathbf{Y}, \mathbf{Z}, \mathbf{S} \mid \mathbf{\Theta})$

$$Q(\mathbf{\Theta}, \mathbf{\Theta}^{old}) = E_{\mathbf{X}, \mathbf{Y}, \mathbf{\Theta}^{old}}\left(\ln P(\mathbf{X}, \mathbf{Y}, \mathbf{Z}, \mathbf{S} \mid \mathbf{\Theta})\right) \quad (23)$$

where, $\mathbf{\Theta}^{old}$ indicates the parameters which are calculated in the previous iteration.

Intuitively, $\mathbf{c}_y$ and $\mathbf{c}_x$ are the expectations of the input $y_t$ and output $x_t$, respectively. They are given by

$$\mathbf{c}_y = \sum_{t=1}^{T} \mathbf{y}_t / T, \quad \mathbf{c}_x = \sum_{t=1}^{T} \mathbf{x}_t / T \quad (24)$$

The optimization of parameters $\mathbf{W}$ and $\mathbf{\Lambda}_\varepsilon$ is different from other model parameters. Actually, both $\mathbf{W}$ and $\mathbf{\Lambda}_\varepsilon$ are determined by parameter $\lambda_i$, and they can be calculated by taking the derivative of Q-function with respect to $\lambda_i$

$$\frac{\partial Q(\mathbf{\Theta}, \mathbf{\Theta}^{old})}{\partial \lambda_i} = \partial\left[\prod_{t=1}^{T} P(\mathbf{z}_t \mid \mathbf{s}_t)\right] \Big/ \partial \lambda_i \quad (25)$$

Set Eq. (25) to be zero, and then we have

$$f(\lambda_i) = \alpha_i^3 \lambda_i^3 + \alpha_i^2 \lambda_i^2 + \alpha_i^1 \lambda_i + \alpha_i^0 = 0 \quad (26)$$

where

$$\begin{aligned}\alpha_i^0 &= -\sum_{t=1}^{T} E_{\mathbf{X},\mathbf{Y},\mathbf{\Theta}^{old}}(s_t^i z_t^i), \quad \alpha_i^2 = -\sum_{t=1}^{T} E_{\mathbf{X},\mathbf{Y},\mathbf{\Theta}^{old}}(s_t^i z_t^i)\\ \alpha_i^1 &= \sum_{t=1}^{T}\left\{E_{\mathbf{X},\mathbf{Y},\mathbf{\Theta}^{old}}\left[(s_t^i)^2 + (z_t^i)^2\right] - 1\right\}, \quad \alpha_i^3 = T\end{aligned} \quad (27)$$

Based on the coefficients given in Eq. (27), and then we have $f(\lambda_i = 0) \leq 0$ and $f(\lambda_i = 1) \geq 0$. Since $f(\lambda_i)$ is a continuous function, there is at least one solution that is larger than 0 and smaller than 1. Based on the result given in Eq. (19), the maximum real root of Eq. (26) which satisfies $\lambda_i \in [0,1]$ is selected as $\lambda_i^{new}$ for the next iteration. After that, the transition matrix $\mathbf{W}$ and covariance matrix $\mathbf{\Lambda}_\varepsilon$ can easily be calculated. Denote $\bar{\mathbf{y}}_t = \mathbf{y}_t - \mathbf{c}_y$ and $\bar{\mathbf{x}}_t = \mathbf{x}_t - \mathbf{c}_x$ for brevity. Then, other parameters in GPMM can be estimated using EM algorithm, and the detailed information is summarized in Table I.

**Lemma 1.** Given $x_t$ and $y_t$, the posterior distribution $s_t$ is a multivariate Gaussian distribution with mean $\boldsymbol{\mu}_{s_t | x_t, y_t}$ and covariance $\boldsymbol{\Xi}_{s_t | x_t, y_t}$ as shown below

$$p(\mathbf{s}_t \mid \mathbf{x}_t, \mathbf{y}_t) \sim N\left(\boldsymbol{\mu}_{s_t | x_t, y_t}, \boldsymbol{\Xi}_{s_t | x_t, y_t}\right) \quad (28)$$

where

$$\begin{aligned}\boldsymbol{\mu}_{s_t | x_t, y_t} &= \boldsymbol{\Xi}_{s_t | x_t, y_t} \times \begin{bmatrix}\mathbf{W}^T\mathbf{U}^T(\mathbf{U}\mathbf{\Lambda}_\varepsilon\mathbf{U}^T + \mathbf{\Lambda}_y)^{-1}\bar{\mathbf{y}}_t\\ +\mathbf{V}^T(\mathbf{\Lambda}_x)^{-1}\bar{\mathbf{x}}_t\end{bmatrix}\\ \boldsymbol{\Xi}_{s_t | x_t, y_t} &= \begin{bmatrix}\mathbf{W}^T\mathbf{U}^T(\mathbf{U}\mathbf{\Lambda}_\varepsilon\mathbf{U}^T + \mathbf{\Lambda}_y)^{-1}\mathbf{U}\mathbf{W}\\ +\mathbf{V}^T(\mathbf{\Lambda}_x)^{-1}\mathbf{V} + \mathbf{I}_r\end{bmatrix}^{-1}\end{aligned} \quad (29)$$

**Proof.** According to Bayesian rule, the posterior distribution of $y_t$ given $s_t$ is provided as below

$$P(\mathbf{y}_t | \mathbf{s}_t) = \int P(\mathbf{y}_t | \mathbf{z}_t) P(\mathbf{z}_t | \mathbf{s}_t) d\mathbf{z}_t$$
$$= N(\mathbf{y}_t | \mathbf{UW}\mathbf{s}_t + \mathbf{c}_y, \mathbf{U}\mathbf{\Lambda}_\varepsilon \mathbf{U}^T + \mathbf{\Lambda}_y) \quad (30)$$

Hence, the joint distribution of $\mathbf{x}_t$ and $\mathbf{y}_t$ when conditioned on the latent variables $\mathbf{s}_t$ can be calculated as below

$$\begin{pmatrix} \mathbf{y}_t \\ \mathbf{x}_t \end{pmatrix} \sim N\left( \begin{bmatrix} \mathbf{UW}\mathbf{s}_t + \mathbf{c}_y \\ \mathbf{V}\mathbf{s}_t + \mathbf{c}_x \end{bmatrix}, \begin{bmatrix} \mathbf{U}\mathbf{\Lambda}_\varepsilon \mathbf{U}^T + \mathbf{\Lambda}_y & \mathbf{0} \\ \mathbf{0} & \mathbf{\Lambda}_x \end{bmatrix} \right) \quad (31)$$

Based on Eq. (31), the joint distribution of $\mathbf{x}_t$ and $\mathbf{y}_t$ is given below when the latent variable is marginalized

$$N\left( \begin{bmatrix} \mathbf{c}_y \\ \mathbf{c}_x \end{bmatrix}, \begin{bmatrix} (\mathbf{UW})(\mathbf{UW})^T + \mathbf{U}\mathbf{\Lambda}_\varepsilon \mathbf{U}^T + \mathbf{\Lambda}_y & \mathbf{UWV}^T \\ \mathbf{V}(\mathbf{UW})^T, & \mathbf{VV}^T + \mathbf{\Lambda}_x \end{bmatrix} \right) \quad (32)$$

From Bayesian rule, the following equation can be obtained
$$P(\mathbf{s}_t | \mathbf{x}_t, \mathbf{y}_t) = P(\mathbf{x}_t, \mathbf{y}_t, \mathbf{s}_t) / P(\mathbf{x}_t, \mathbf{y}_t) \quad (33)$$

Since $P(\mathbf{x}_t, \mathbf{y}_t)$ is irrelevant to $\mathbf{s}_t$, it can be omitted in order to derive the posterior distribution $P(\mathbf{s}_t | \mathbf{x}_t, \mathbf{y}_t)$. Hence, Eq. (33) can be further recast as following
$$P(\mathbf{s}_t | \mathbf{x}_t, \mathbf{y}_t) \propto P(\mathbf{x}_t, \mathbf{y}_t, \mathbf{s}_t) \quad (34)$$
where
$$P(\mathbf{s}_t | \mathbf{x}_t, \mathbf{y}_t) \propto P(\mathbf{x}_t | \mathbf{s}_t) \times P(\mathbf{y}_t | \mathbf{s}_t) \times P(\mathbf{s}_t)$$
$$= \exp\left\{ -\frac{1}{2}(\bar{\mathbf{y}}_t - \mathbf{UW}\mathbf{s}_t)^T (\mathbf{U}\mathbf{\Lambda}_\varepsilon \mathbf{U}^T + \mathbf{\Lambda}_y)^{-1} (\bar{\mathbf{y}}_t - \mathbf{UW}\mathbf{s}_t) \right\}$$
$$\times \exp\left\{ -\frac{1}{2}(\bar{\mathbf{x}}_t - \mathbf{V}\mathbf{s}_t)^T (\mathbf{\Lambda}_x)^{-1} (\bar{\mathbf{x}}_t - \mathbf{V}\mathbf{s}_t) \right\} \times \exp\left\{ -\frac{1}{2}\mathbf{s}_t^T \mathbf{s}_t \right\} \quad (35)$$

Recast the exponents of the above equation as a quadratic function in $\mathbf{s}_t$ and drop the other constant terms. Then Eq. (35) can be transformed as

$$\exp\left\{ \begin{array}{l} -\frac{1}{2}\mathbf{s}_t^T \left[ \mathbf{W}^T \mathbf{U}^T \mathbf{\Omega}^{-1} \mathbf{UW} + \mathbf{V}^T (\mathbf{\Lambda}_x)^{-1} \mathbf{V} + \mathbf{I}_r \right] \mathbf{s}_t \\ +\mathbf{s}_t^T \left[ \mathbf{W}^T \mathbf{U}^T \mathbf{\Omega}^{-1} \bar{\mathbf{y}}_t + \mathbf{V}^T (\mathbf{\Lambda}_x)^{-1} \bar{\mathbf{x}}_t \right] \end{array} \right\} \quad (36)$$

where, $\mathbf{\Omega} = \mathbf{U}\mathbf{\Lambda}_\varepsilon \mathbf{U}^T + \mathbf{\Lambda}_y$.

It is easy to find that the expectation $\mu_{\mathbf{s}_t | \mathbf{x}_t, \mathbf{y}_t}$ and covariance $\Xi_{\mathbf{s}_t | \mathbf{x}_t, \mathbf{y}_t}$ are identical to the ones shown in Eq. (29).

**Lemma 2.** Given $\mathbf{x}_t$ and $\mathbf{y}_t$, the posterior distribution $\mathbf{z}_t$ is a multivariate Gaussian distribution whose detailed expression is given by
$$p(\mathbf{z}_t | \mathbf{x}_t, \mathbf{y}_t) \sim N\left( \mu_{\mathbf{z}_t | \mathbf{x}_t, \mathbf{y}_t}, \Xi_{\mathbf{z}_t | \mathbf{x}_t, \mathbf{y}_t} \right) \quad (37)$$

The detailed information of expectation $\mu_{\mathbf{z}_t | \mathbf{x}_t, \mathbf{y}_t}$ and covariance $\Xi_{\mathbf{z}_t | \mathbf{x}_t, \mathbf{y}_t}$ is given as below

$$\mu_{\mathbf{z}_t | \mathbf{x}_t, \mathbf{y}_t} = \Xi_{\mathbf{z}_t | \mathbf{x}_t, \mathbf{y}_t} \times \left[ \mathbf{M}^T (\mathbf{N})^{-1} \bar{\mathbf{x}}_t + \mathbf{U}^T (\mathbf{\Lambda}_y)^{-1} \bar{\mathbf{y}}_t \right]$$
$$\Xi_{\mathbf{z}_t | \mathbf{x}_t, \mathbf{y}_t}^{-1} = \mathbf{M}^T (\mathbf{N})^{-1} \mathbf{M} + \mathbf{U}^T (\mathbf{\Lambda}_y)^{-1} \mathbf{U} + (\mathbf{\Lambda}_\varepsilon + \mathbf{WW}^T)^{-1} \quad (38)$$

where
$$\mathbf{M} = \mathbf{V}\left[ \mathbf{I}_r + \mathbf{W}^T (\mathbf{\Lambda}_\varepsilon)^{-1} \mathbf{W} \right]^{-1} \left[ \mathbf{W}^T (\mathbf{\Lambda}_\varepsilon)^{-1} \right]$$
$$\mathbf{N} = \mathbf{\Lambda}_x + \mathbf{V}\left[ \mathbf{I}_r + \mathbf{W}^T (\mathbf{\Lambda}_\varepsilon)^{-1} \mathbf{W} \right] \mathbf{V}^T \quad (39)$$

**Proof.** From Bayesian rule, the posterior distribution of $\mathbf{x}_t$ given $\mathbf{z}_t$ is provided by

$$P(\mathbf{x}_t | \mathbf{z}_t) = \int P(\mathbf{x}_t | \mathbf{s}_t) P(\mathbf{s}_t | \mathbf{z}_t) d\mathbf{s}_t \quad (40)$$
where, $P(\mathbf{z}_t | \mathbf{s}_t) = N(\mathbf{z}_t | \mathbf{W}\mathbf{s}_t, \mathbf{\Lambda}_\varepsilon)$, $P(\mathbf{s}_t) = N(\mathbf{s}_t | \mathbf{0}, \mathbf{I}_r)$.

Based on the above equation, we have
$$P(\mathbf{s}_t | \mathbf{z}_t) = N\left(\mathbf{s}_t | \left[ \mathbf{I}_r + \mathbf{W}^T (\mathbf{\Lambda}_\varepsilon)^{-1} \mathbf{W} \right]^{-1} \left[ \mathbf{W}^T (\mathbf{\Lambda}_\varepsilon)^{-1} \right] \mathbf{z}_t, \right.$$
$$\left. \left[ \mathbf{I}_r + \mathbf{W}^T (\mathbf{\Lambda}_\varepsilon)^{-1} \mathbf{W} \right]^{-1} \right) \quad (41)$$

Substituting Eq. (41) in Eq. (40), and $P(\mathbf{x}_t | \mathbf{z}_t)$ can be further written as below
$$P(\mathbf{x}_t | \mathbf{z}_t) = N(\mathbf{x}_t | \mathbf{M}\mathbf{z}_t + \mathbf{c}_x, \mathbf{N}) \quad (42)$$
where, $\mathbf{M}$ and $\mathbf{N}$ are given in Eq. (39).

According to Bayesian rule, we have the following equation
$$P(\mathbf{z}_t | \mathbf{x}_t, \mathbf{y}_t) = P(\mathbf{x}_t, \mathbf{y}_t, \mathbf{z}_t) / P(\mathbf{x}_t, \mathbf{y}_t) \quad (43)$$

Similar to Eq. (33), Eq. (43) can be further recast as below
$$P(\mathbf{z}_t | \mathbf{x}_t, \mathbf{y}_t) \propto P(\mathbf{x}_t | \mathbf{z}_t) \times P(\mathbf{y}_t | \mathbf{z}_t) \times P(\mathbf{z}_t) \quad (44)$$

The exponents of the above equation can be rewritten as a quadratic function in $\mathbf{z}_t$

$$-\frac{1}{2}\mathbf{z}_t^T \left[ \mathbf{M}^T (\mathbf{N})^{-1} \mathbf{M} + \mathbf{U}^T (\mathbf{\Lambda}_y)^{-1} \mathbf{U} + (\mathbf{\Lambda}_\varepsilon + \mathbf{WW}^T)^{-1} \right] \mathbf{z}_t$$
$$+\mathbf{z}_t^T \left[ \mathbf{M}^T (\mathbf{N})^{-1} \bar{\mathbf{x}}_t + \mathbf{U}^T (\mathbf{\Lambda}_y)^{-1} \bar{\mathbf{y}}_t \right] \quad (45)$$

where, the expressions of $\mu_{\mathbf{z}_t | \mathbf{x}_t, \mathbf{y}_t}$ and $\Xi_{\mathbf{z}_t | \mathbf{x}_t, \mathbf{y}_t}^{-1}$ are identical to the ones shown in Eqs. (38) and (39).

### 3.3 Solution of GPMM with Sequential Data

When developing GPMM with sequential data, the optimization problem transforms into multiple Markov chains with identical parameters. The graphical structure of GPMM with sequential data is depicted in Fig. 2 for illustration. As shown in this figure, the latent variable $\mathbf{s}_{t+\tau}$ is conditioned on the preceding $\mathbf{s}_t$, and the system output $\mathbf{x}_t$ is conditioned on its corresponding latent variable $\mathbf{s}_t$. For this case, the unknown parameters are defined as $\Theta = \{\mathbf{V}, \mathbf{W}, \mathbf{\Lambda}_x, \mathbf{\Lambda}_\varepsilon\}$.

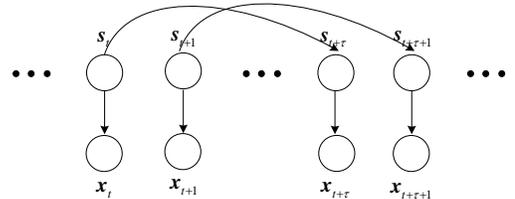

Fig. 2. The graphical structure of GPMM with sequential data.

For this case, the training data is decomposed into $\tau$ subsequences $\mathbf{X}_k = \{\mathbf{x}_1^k, \mathbf{x}_2^k, \cdots \mathbf{x}_{T_k}^k\}$ $k = 1, 2, \cdots \tau$. By this means, the complete data likelihood of GPMM is given as

$$P = \prod_{k=1}^{\tau} \left\{ P(\mathbf{s}_1^k) \times \prod_{t=2}^{T_k} P(\mathbf{s}_t^k | \mathbf{s}_{t-1}^k) \times \prod_{t=1}^{T_k} P(\mathbf{x}_t^k | \mathbf{s}_t^k) \right\} \quad (46)$$

where
$$P(\mathbf{s}_t^k | \mathbf{s}_{t-1}^k) = N(\mathbf{s}_t^k | \mathbf{W}\mathbf{s}_{t-1}^k, \mathbf{\Lambda}_\varepsilon)$$
$$P(\mathbf{x}_t^k | \mathbf{s}_t^k) = N(\mathbf{x}_t^k | \mathbf{V}\mathbf{s}_t^k + \mathbf{c}_x, \mathbf{\Lambda}_x) \quad (47)$$
$$P(\mathbf{s}_1^k) = N(\mathbf{0}, \mathbf{I}_r) \quad k = 1, 2, \cdots, \tau$$

Calculate the logarithm of Eq. (46), and then denote the Q-function as below

$$Q(\Theta, \Theta^{old}) = E_{\mathbf{X},\mathbf{Y},\Theta^{old}}\left(\ln P(\mathbf{X},\mathbf{S}\mid \Theta)\right) \quad (48)$$

where, $\Theta^{old}$ indicates the parameters which are calculated in the last iteration.

In the M-Step, the Q-function should be maximized with respect to parameters $\Theta$. Since the M-steps in GPMM with random and sequential data are almost the same, the procedure is omitted in this study to simplify the presentation.

In the E-Step, the expectations of the posterior distributions are calculated. The solution of E-Step is solved using the forward-backward algorithm, which has been introduced by Bishop (2006). For brevity, the expectations based on $\mathbf{X}_k$ are provided as below, and the expectations that are needed in M-Step can be regarded as the weighted average of each subsequence.

$$E_{\mathbf{X},\Theta^{old}}(s_t^k) = \hat{\boldsymbol{\mu}}_t^k, \quad E_{\mathbf{X},\Theta^{old}}\left[(s_t^k)(s_t^k)^T\right] = \hat{\Upsilon}_t^k + (\hat{\boldsymbol{\mu}}_t^k)(\hat{\boldsymbol{\mu}}_t^k)^T$$
$$E_{\mathbf{X},\Theta^{old}}\left[(s_t^k)(s_{t-1}^k)^T\right] = \mathbf{J}_{t-1}^k \hat{\Upsilon}_t^k + (\hat{\boldsymbol{\mu}}_t^k)(\hat{\boldsymbol{\mu}}_{t-1}^k)^T \quad (49)$$

Taking the k-th subsequence $\mathbf{X}_k = \{x_1^k, x_2^k, \cdots x_{T_k}^k\}$ as an example, the posterior distribution $P(s_t^k \mid x_1^k, x_2^k, \cdots x_t^k, \Theta^{old})$ is calculated in the forward recursion. Actually, a Kalman smoothing problem is involved in the inference of latent variables $\mathbf{S}_k = \{s_1^k, s_2^k, \cdots s_{T_k}^k\}$ given samples $\mathbf{X}_k$.

$$\boldsymbol{\mu}_t^k = \mathbf{W}\boldsymbol{\mu}_{t-1}^k + \mathbf{K}_t^k(\bar{x}_t^k - \mathbf{V}\mathbf{W}\boldsymbol{\mu}_{t-1}^k)$$
$$\Upsilon_t^k = (\mathbf{I}_r - \mathbf{K}_t^k \mathbf{V})(\Lambda_\varepsilon + \mathbf{W}\Upsilon_{t-1}^k \mathbf{W}^T)$$
$$\mathbf{K}_t^k = \left(\Lambda_\varepsilon + \mathbf{W}\Upsilon_{t-1}^k \mathbf{W}^T\right)\mathbf{V}^T \quad (50)$$
$$\times\left[\mathbf{V}\left(\Lambda_\varepsilon + \mathbf{W}\Upsilon_{t-1}^k \mathbf{W}^T\right)\mathbf{V}^T + \Lambda_x\right]^{-1}$$

where

$$\boldsymbol{\mu}_1^k = \mathbf{K}_1^k \bar{x}_1^k, \quad \Upsilon_1^k = (\mathbf{I}_r - \mathbf{K}_1^k \mathbf{V})\left(\Lambda_\varepsilon + \mathbf{W}\mathbf{W}^T\right)$$
$$\mathbf{K}_1^k = \left(\Lambda_\varepsilon + \mathbf{W}\mathbf{W}^T\right)\mathbf{V}^T\left[\mathbf{V}\left(\Lambda_\varepsilon + \mathbf{W}\mathbf{W}^T\right)\mathbf{V}^T + \Lambda_x\right]^{-1} \quad (51)$$

Using backward recursion, the parameters of the posterior distribution $P(\mathbf{S}_k \mid \mathbf{X}_k, \Theta^{old})$ can be calculated by

$$\hat{\boldsymbol{\mu}}_t^k = \boldsymbol{\mu}_t^k + \mathbf{J}_t^k(\hat{\boldsymbol{\mu}}_{t+1}^k + \mathbf{W}\boldsymbol{\mu}_t^k)$$
$$\hat{\Upsilon}_t^k = \Upsilon_t^k + \mathbf{J}_t^k(\hat{\Upsilon}_{t+1}^k - \mathbf{W}\Upsilon_t^k \mathbf{W}^T - \Lambda_\varepsilon)(\mathbf{J}_t^k)^T \quad (52)$$
$$\mathbf{J}_t^k = \Upsilon_t^k \mathbf{W}^T(\mathbf{W}\Upsilon_t^k \mathbf{W}^T + \Lambda_\varepsilon)^{-1}$$

with initializations $\hat{\boldsymbol{\mu}}_T^k = \boldsymbol{\mu}_T^k$, $\hat{\Upsilon}_T^k = \Upsilon_T^k$.

In GPMM model with sequential data, all the samples in subsequence $\mathbf{X}_k$ should be considered (i.e., $P(s_k^t \mid \mathbf{X}_k)$). This is totally different from the model with random data, which only take the current samples $(x_t, y_t)$ into consideration.

## 4 GPMM based Monitoring Strategy

### 4.1 Monitoring Statistics in GPMM with Random Data
#### 4.1.1 Monitoring statistics of latent variables
**Theorem 1.** Assume that the observations of the system are accurately defined by GPMM with exact parameters. Whether $s_t$ and $z_t$ exceed their normal operation regions with $(1-\alpha)$ confidence level can be verified by checking if the following inequalities violate. In general, $\alpha$ is set to 1% or 5% in most conventional monitoring strategies.

$$T_s^{Ran} = \left(\boldsymbol{\mu}_{s_t\mid x_t, y_t}\right)^T\left(\mathbf{I}_r - \Xi_{s_t\mid x_t, y_t}\right)^{-1}\left(\boldsymbol{\mu}_{s_t\mid x_t, y_t}\right) \leq \chi^2_{(1-\alpha; \theta_s)} \quad (53)$$

$$T_z^{Ran} = \left(\boldsymbol{\mu}_{z_t\mid x_t, y_t}\right)^T\left(\mathbf{I}_r - \Xi_{z_t\mid x_t, y_t}\right)^{-1}\left(\boldsymbol{\mu}_{z_t\mid x_t, y_t}\right) \leq \chi^2_{(1-\alpha; \theta_z)} \quad (54)$$

where, the detailed information of $\boldsymbol{\mu}_{s_t\mid x_t, y_t}$, $\boldsymbol{\mu}_{z_t\mid x_t, y_t}$, $\Xi_{s_t\mid x_t, y_t}$, and $\Xi_{z_t\mid x_t, y_t}$ have been provided in Eqs. (29) and (38).

**Proof.** From Eqs. (29) and (38), it is easy to find that $\boldsymbol{\mu}_{s_t\mid x_t, y_t}$ and $\boldsymbol{\mu}_{z_t\mid x_t, y_t}$ are the linear combinations of the observations $x_t$ and $y_t$. Besides, $\Xi_{s_t\mid x_t, y_t}$ and $\Xi_{z_t\mid x_t, y_t}$ have no direct relationship with these observations. Hence, it is sufficient to monitor only $\boldsymbol{\mu}_{s_t\mid x_t, y_t}$ and $\boldsymbol{\mu}_{z_t\mid x_t, y_t}$.

In the following, we will demonstrate that the statistic $T_s^{Ran}$ follows a chi-squared distribution with $\theta_s$-degrees of freedom. This is equivalent to prove that the expectation and covariance of $\boldsymbol{\mu}_{s_t\mid x_t, y_t}$ are zero and $\mathbf{I}_r - \Xi_{s_t\mid x_t, y_t}$, respectively.

Since $\boldsymbol{\mu}_{s_t\mid x_t, y_t}$ is a linear combination of the observations, it follows a multivariate Gaussian distribution.

$$E(\boldsymbol{\mu}_{s_t\mid x_t, y_t}) = \Xi_{s_t\mid x_t, y_t}\begin{bmatrix}\mathbf{W}^T\mathbf{U}^T(\mathbf{U}\Lambda_\varepsilon\mathbf{U}^T + \Lambda_y)^{-1}E(\bar{y}_t) \\ +\mathbf{V}^T(\Lambda_x)^{-1}E(\bar{x}_t)\end{bmatrix} = 0 \quad (55)$$

Based on Eq. (55), the covariance of $\boldsymbol{\mu}_{s_t\mid x_t, y_t}$ can be calculated using the following equation

$$Cov(\boldsymbol{\mu}_{s_t\mid x_t, y_t}) = E(\boldsymbol{\mu}_{s_t\mid x_t, y_t}\boldsymbol{\mu}_{s_t\mid x_t, y_t}^T) =$$
$$\Xi_{s_t\mid x_t, y_t}\mathbf{W}^T\mathbf{U}^T(\mathbf{U}\Lambda_\varepsilon\mathbf{U}^T + \Lambda_y)^{-1}E(\bar{y}_t\bar{y}_t^T)(\mathbf{U}\Lambda_\varepsilon\mathbf{U}^T + \Lambda_y)^{-1}$$
$$\times\mathbf{U}\mathbf{W}\Xi_{s_t\mid x_t, y_t} + \Xi_{s_t\mid x_t, y_t}\mathbf{V}^T(\Lambda_x)^{-1}E(\bar{x}_t\bar{x}_t^T)(\Lambda_x)^{-1}\mathbf{V}\Xi_{s_t\mid x_t, y_t} \quad (56)$$
$$+\Xi_{s_t\mid x_t, y_t}\mathbf{W}^T\mathbf{U}^T(\mathbf{U}\Lambda_\varepsilon\mathbf{U}^T + \Lambda_y)^{-1}E(\bar{y}_t\bar{x}_t^T)(\Lambda_x)^{-1}\mathbf{V}\Xi_{s_t\mid x_t, y_t}$$
$$+\Xi_{s_t\mid x_t, y_t}\mathbf{V}^T(\Lambda_x)^{-1}E(\bar{x}_t\bar{y}_t^T)(\mathbf{U}\Lambda_\varepsilon\mathbf{U}^T + \Lambda_y)^{-1}\mathbf{U}\mathbf{W}\Xi_{s_t\mid x_t, y_t}$$

Substituting Eq. (17) in the above equation, and it can be further simplified to a compact form

$$= \Xi_{s_t\mid x_t, y_t}\left\{(\Xi_{s_t\mid x_t, y_t})^{-1} - \mathbf{I}_r\right\} = \mathbf{I}_r - \Xi_{s_t\mid x_t, y_t} \quad (57)$$

Since the dimension of $\boldsymbol{\mu}_{s_t\mid x_t, y_t}$ is $r$, the statistic $T_s^{Ran}$ follows the chi-square distribution with $\theta_s$ degrees (i.e., $\theta_s \leq r$) of freedom. The demonstration procedure of $T_z^{Ran}$ is similar to that of $T_s^{Ran}$, and thus it is omitted for brevity.

#### 4.1.2 Monitoring statistics of model residual
The covariance matrices of $s_t$ and $z_t$ are identity matrices, and they are different from that of $\boldsymbol{\mu}_{s_t\mid x_t, y_t}$ and $\boldsymbol{\mu}_{z_t\mid x_t, y_t}$. Hence, $\boldsymbol{\mu}_{s_t\mid x_t, y_t}$ and $\boldsymbol{\mu}_{z_t\mid x_t, y_t}$ may not be used as the optimal reconstructions of GPMM. Based on the relationship $z_t = \mathbf{W}s_t + \varepsilon$, we have the following equation

$$\bar{y}_t - \mathbf{U}\mathbf{W}s_t = e_y + \mathbf{U}\varepsilon \quad (58)$$

Using Eqs. (17) and (58), the optimal reconstruction can be obtained by solving the following equation

$$\min_{s_t}(\bar{y}_t - \mathbf{U}\mathbf{W}s_t)^T(\mathbf{U}\Lambda_\varepsilon\mathbf{U}^T + \Lambda_y)^{-1}(\bar{y}_t - \mathbf{U}\mathbf{W}s_t)$$
$$+(\bar{x}_t - \mathbf{V}s_t)^T(\Lambda_x)^{-1}(\bar{x}_t - \mathbf{V}s_t) \quad (59)$$

Based on the solution of the above optimization problem, a monitoring statistic can be developed to check how well the observations can be predicted. The optimization problem of Eq. (59) can be solved by setting its derivative to zero with respect to latent variable $s_t$. The solution is given by

$$s_t = \left[ \mathbf{W}^T \mathbf{U}^T (\mathbf{U}\mathbf{\Lambda}_\varepsilon \mathbf{U}^T + \mathbf{\Lambda}_y)^{-1} \mathbf{U}\mathbf{W} + \mathbf{V}^T (\mathbf{\Lambda}_x)^{-1} \mathbf{V} \right]^{-1}$$
$$\times \begin{pmatrix} \mathbf{U}\mathbf{W} \\ \mathbf{V} \end{pmatrix}^T \begin{pmatrix} (\mathbf{U}\mathbf{\Lambda}_\varepsilon \mathbf{U}^T + \mathbf{\Lambda}_y)^{-1} & \\ & (\mathbf{\Lambda}_x)^{-1} \end{pmatrix} \begin{pmatrix} \bar{y}_t \\ \bar{x}_t \end{pmatrix} \quad (60)$$

**Theorem 2.** Assume that the distribution of the observation has been accurately defined by GPMM with the exact parameters. Then, the value of Eq. (60) follows a chi-square distribution with $\kappa_d$ (i.e., $\kappa_d \leq p+q$) degrees of freedom.

**Proof.** Substituting Eq. (60) in Eq. (59), and then Eq. (59) can be recast as a compact form as below

$$\begin{pmatrix} \bar{y}_t \\ \bar{x}_t \end{pmatrix}^T \mho^{-1/2} \left\{ \mathbf{I}_{p+q} - \wp(\wp^T \wp)^{-1} \wp^T \right\} \mho^{-1/2} \begin{pmatrix} \bar{y}_t \\ \bar{x}_t \end{pmatrix} \quad (61)$$

where

$$\mho^{-1} = \begin{pmatrix} (\mathbf{U}\mathbf{\Lambda}_\varepsilon \mathbf{U}^T + \mathbf{\Lambda}_y)^{-1} & \\ & (\mathbf{\Lambda}_x)^{-1} \end{pmatrix}, \quad \wp = \mho^{-1/2} \begin{pmatrix} \mathbf{U}\mathbf{W} \\ \mathbf{V} \end{pmatrix} \quad (62)$$

Define $\mathbf{A} = \mho^{-1/2} \left\{ \mathbf{I}_{p+q} - \wp(\wp^T \wp)^{-1} \wp^T \right\} \mho^{-1/2}$, and the term $\left\{ \mathbf{I}_{p+q} - \wp(\wp^T \wp)^{-1} \wp^T \right\}$ is an idempotent matrix. According to Eq. (32), the whitening transformation matrix is given by

$$\aleph = \begin{bmatrix} (\mathbf{U}\mathbf{W})(\mathbf{U}\mathbf{W})^T + \mathbf{U}\mathbf{\Lambda}_\varepsilon \mathbf{U}^T + \mathbf{\Lambda}_y & \mathbf{U}\mathbf{W}\mathbf{V}^T \\ \mathbf{V}(\mathbf{U}\mathbf{W})^T & \mathbf{V}\mathbf{V}^T + \mathbf{\Lambda}_x \end{bmatrix} \quad (63)$$

Set $\ell_t = \begin{pmatrix} \bar{y}_t \\ \bar{x}_t \end{pmatrix}$, and it is obvious that the covariance matrix of $\hat{\ell}_t = \aleph^{-1/2} \ell_t$ is an identity matrix.

$$(\ell_t)^T \mathbf{A}(\ell_t) = (\hat{\ell}_t)^T \aleph^{1/2} \mathbf{A} \aleph^{1/2} (\hat{\ell}_t) = (\hat{\ell}_t)^T \mathfrak{A}(\hat{\ell}_t) \quad (64)$$

Since $\mathfrak{A}$ is a symmetric matrix, it can be decomposed using SVD as below

$$(\hat{\ell}_t)^T \mathfrak{A}(\hat{\ell}_t) = (\mathfrak{I}\hat{\ell}_t)^T \mathbf{\Omega}(\mathfrak{I}\hat{\ell}_t) = \eta^T \mathbf{\Omega} \eta \quad (65)$$

where, $\mathbf{\Omega}$ is a diagonal matrix which is composed by the eigenvalues of $\mathfrak{A}$. $\mathfrak{I}$ consists of the eigenvalues of $\mathfrak{A}$ and it satisfies $\mathfrak{I}^T \mathfrak{I} = \mathbf{I}_{p+q}$.

It is easy to demonstrate that $\eta \sim N(0, \mathbf{I}_{p+q})$, and thus the proof of **Theorem 2** is equivalent to the prove that $\mathbf{\Omega}$ consists of only 0 and 1. As we know, the eigenvalues of an idempotent matrix are 0 and 1. If $\mathfrak{A}$ is an idempotent matrix, $(\ell_t)^T \mathbf{A}(\ell_t)$ is a non-negative sum of chi-square random variables. Hence, the proof of **Theorem 2** reduces to be

$$\mathfrak{A}\mathfrak{A} = \aleph^{1/2} \mathbf{A} \aleph \mathbf{A} \aleph^{1/2} = \mathfrak{A} \quad (66)$$

which is equal to show

$$\mathbf{A} \aleph \mathbf{A} = \mathbf{A} \quad (67)$$

Substituting Eq. (63) and the express of $\mathbf{A}$ in Eq. (67), and then we have

$$\mathbf{A} \aleph \mathbf{A} = \mho^{-1/2} \left\{ \mathbf{I}_{p+q} - \wp(\wp^T \wp)^{-1} \wp^T \right\} \mho^{-1/2} = \mathbf{A} \quad (68)$$

An idempotent matrix is nonsingular unless it is an identity matrix. Since $\wp(\wp^T \wp)^{-1} \wp^T$ is not a zero matrix, $\mathfrak{A}$ is rank deficient and thus $\kappa_d \leq p+q$, where $\kappa_d$ indicates the rank of $\mathfrak{A}$. Therefore, the value of Eq. (59) follows chi-square distribution with $\kappa_d$ degrees of freedom. Similar to the monitoring policy of latent variable, the operation status of the system can also be checked by verifying if the following inequality violates

$$Q_{Ran} = (\ell_t)^T \mathbf{A}(\ell_t) \leq \chi^{-2}_{(1-\alpha, \kappa_d)} \quad (69)$$

### 4.1.3 Other possible monitoring statistics

Additional monitoring statistics can also be designed to reflect other aspects of the system, such as properties related to latent variables $s_t$ and $z_t$ which are inferred from a partial set of observations.

Similar to **Lemma 1**, the posterior distribution of $s_t$ when given only observation $x_t$ is depicted below

$$p(s_t | x_t) \sim N(\mu_{s_t | x_t}, \Xi_{s_t | x_t}) \quad (70)$$

where

$$\mu_{s_t | x_t} = \Xi_{s_t | x_t} \mathbf{V}^T (\mathbf{\Lambda}_x)^{-1} \bar{x}_t, \quad \Xi_{s_t | x_t} = \left[ \mathbf{V}^T (\mathbf{\Lambda}_x)^{-1} \mathbf{V} + \mathbf{I}_r \right]^{-1} \quad (71)$$

It is noted that the results given in Eq. (29) are equivalent to the solutions in Eq. (71) when setting $\mathbf{U} = 0$. Based on Eq. (71), the monitoring statistic $T_s^p$ of latent variables $s_t$ when given only observation $x_t$ is designed through Eq. (72). The distribution of this statistic can be derived using a similar procedure of the proof in **Theorem 1**.

$$T_s^p = \mu_{s_t | x_t}^T \left[ \mathbf{I}_r - \Xi_{s_t | x_t} \right]^{-1} \mu_{s_t | x_t} \leq \chi^{-2}_{(1-\alpha, \theta_s^p)} \quad (72)$$

Furthermore, the monitoring statistic $T_z^p$ of latent variable $z_t$ when giving only observation $y_t$ can be derived by

$$T_z^p = \mu_{z_t | y_t}^T \left[ \mathbf{I}_r - \Xi_{z_t | y_t} \right]^{-1} \mu_{z_t | y_t} \leq \chi^{-2}_{(1-\alpha, \theta_z^p)} \quad (73)$$

where

$$\mu_{z_t | y_t} = \Xi_{z_t | x_t, y_t} \mathbf{U}^T (\mathbf{\Lambda}_y)^{-1} \bar{y}_t, \quad \Xi_{z_t | y_t} = \left[ \mathbf{U}^T (\mathbf{\Lambda}_y)^{-1} \mathbf{U} + \mathbf{I}_r \right]^{-1} \quad (74)$$

The results in Eq. (74) can be directly derived from Eq. (38) by setting $\mathbf{V} = 0$ and $\mathbf{\Lambda}_\varepsilon + \mathbf{W}\mathbf{W}^T = \mathbf{I}_r$. The distribution of statistic $T_z^p$ can be proved using procedure which is similar to that of statistic $T_s^p$.

### 4.2 Monitoring Statistics in GPMM with Sequential Data

#### 4.2.1 Monitoring statistics of latent variable

Similar to Section 4.1.1, the latent variable $s_t$ is substituted by $E_{\mathbf{X}, \Theta^{old}}(s_t^k) = \hat{\mu}_t^k$ since it cannot be measured directly. The expectation of $\hat{\mu}_t^k$ can be calculated by

$$E(\hat{\mu}_t^k) = (\mathbf{I}_r + \mathbf{J}_t^k \mathbf{W}) E(\mu_t^k) + \mathbf{J}_t^k E(\hat{\mu}_{t+1}^k) = \mathbf{0} \quad (75)$$

The covariance of $\hat{\mu}_t^k$ can then be calculated using equation which is provided below

$$E\left[ (\hat{\mu}_t^k)(\hat{\mu}_t^k)^T \right] = (\mathbf{I}_r + \mathbf{J}_t^k \mathbf{W}) E\left[ (\mu_t^k)(\mu_t^k)^T \right] (\mathbf{I}_r + \mathbf{J}_t^k \mathbf{W})^T$$
$$+ 2(\mathbf{I}_r + \mathbf{J}_t^k \mathbf{W}) E\left[ (\mu_t^k)(\hat{\mu}_{t+1}^k)^T \right] (\mathbf{J}_t^k)^T \quad (76)$$
$$+ \mathbf{J}_t^k E\left[ (\hat{\mu}_{t+1}^k)(\hat{\mu}_{t+1}^k)^T \right] (\mathbf{J}_t^k)^T$$

From Eq. (76), it is easy to find the recursive relationship between $E\left[(\hat{\boldsymbol{\mu}}_t^k)(\hat{\boldsymbol{\mu}}_t^k)^T\right]$ and $E\left[(\hat{\boldsymbol{\mu}}_{t+1}^k)(\hat{\boldsymbol{\mu}}_{t+1}^k)^T\right]$. Besides, the relationships between $E\left[(\boldsymbol{\mu}_t^k)(\boldsymbol{\mu}_t^k)^T\right]$ and $E\left[(\boldsymbol{\mu}_t^k)(\hat{\boldsymbol{\mu}}_{t+1}^k)^T\right]$ can also be inferred using Eqs. (50) and (52). Based on the recursive relationship, the analytic solution of Eq. (76) can be obtained. Since the solution is too complex and cannot be simplified to a compact form, it is suggested to approximate $E\left[(\hat{\boldsymbol{\mu}}_t^k)(\hat{\boldsymbol{\mu}}_t^k)^T\right]$ using a numerical method. The monitoring policy for $s_t$ in sequential data is given as below

$$T_{seq} = \left(\hat{\boldsymbol{\mu}}_t^k\right)^T \left\{E\left[(\hat{\boldsymbol{\mu}}_t^k)(\hat{\boldsymbol{\mu}}_t^k)^T\right]\right\}^{-1} \left(\hat{\boldsymbol{\mu}}_t^k\right) \leq \chi^2_{(1-\alpha,\vartheta_{seq})} \quad (77)$$

*4.2.2 Monitoring statistics of model residual*

According to Eq. (20), the GPMM can also be formulated for sequential data as following

$$\begin{aligned}
\boldsymbol{x}_{t+\tau} &= \mathbf{V}\boldsymbol{s}_{t+\tau} + \boldsymbol{c}_x + \boldsymbol{e}_x, \quad \boldsymbol{e}_x \sim N(0, \boldsymbol{\Lambda}_x) \\
\boldsymbol{x}_t &= \mathbf{V}\boldsymbol{s}_t + \boldsymbol{c}_x + \boldsymbol{e}_x, \quad \boldsymbol{e}_x \sim N(0, \boldsymbol{\Lambda}_x) \\
\boldsymbol{s}_{t+\tau} &= \mathbf{W}\boldsymbol{s}_t + \boldsymbol{\varepsilon}, \quad \boldsymbol{\varepsilon} \sim N(0, \boldsymbol{\Lambda}_\varepsilon)
\end{aligned} \quad (78)$$

Substituting the third equation in the first one, and we obtain the following relationship

$$\bar{\boldsymbol{x}}_{t+\tau} = \mathbf{V}\mathbf{W}\boldsymbol{s}_t + \mathbf{V}\boldsymbol{\varepsilon} + \boldsymbol{e}_x \quad (79)$$

Based on the first two equations in Eq. (78), the following optimization problem can be formulated to obtain the optimal reconstruction solution

$$\begin{aligned}
\min_{\boldsymbol{s}_t} \ & (\bar{\boldsymbol{x}}_{t+\tau} - \mathbf{V}\mathbf{W}\boldsymbol{s}_t)^T (\mathbf{V}\boldsymbol{\Lambda}_\varepsilon \mathbf{V}^T + \boldsymbol{\Lambda}_x)^{-1} (\bar{\boldsymbol{x}}_{t+\tau} - \mathbf{V}\mathbf{W}\boldsymbol{s}_t) \\
& + (\bar{\boldsymbol{x}}_t - \mathbf{V}\boldsymbol{s}_t)^T (\boldsymbol{\Lambda}_x)^{-1} (\bar{\boldsymbol{x}}_t - \mathbf{V}\boldsymbol{s}_t)
\end{aligned} \quad (80)$$

The optimal solution of Eq. (80) is given by

$$\begin{aligned}
\boldsymbol{s}_t = & \left[\mathbf{W}^T\mathbf{V}^T(\mathbf{V}\boldsymbol{\Lambda}_\varepsilon\mathbf{V}^T + \boldsymbol{\Lambda}_x)^{-1}\mathbf{V}\mathbf{W} + \mathbf{V}^T(\boldsymbol{\Lambda}_x)^{-1}\mathbf{V}\right]^{-1} \\
& \times \left[\mathbf{W}^T\mathbf{V}^T(\mathbf{V}\boldsymbol{\Lambda}_\varepsilon\mathbf{V}^T + \boldsymbol{\Lambda}_x)^{-1}(\bar{\boldsymbol{x}}_{t+\tau}) + \mathbf{V}^T(\boldsymbol{\Lambda}_x)^{-1}(\bar{\boldsymbol{x}}_t)\right]
\end{aligned} \quad (81)$$

**Theorem 3.** Assume that the distribution of the observations has been accurately defined by the GPMM with true parameters. Then, the value of Eq. (80) follows chi-square distribution. Besides, the degrees of freedom of Eq. (80) satisfies $\kappa_s < 2q$.

**Proof.** Substituting Eq. (81) in Eq. (80), and then Eq. (80) can be simplified as below

$$(\boldsymbol{\gamma}_t)^T \boldsymbol{\Phi}^{-1/2} \left\{\mathbf{I}_{2q} - \boldsymbol{\rho}(\boldsymbol{\rho}^T\boldsymbol{\rho})^{-1}\boldsymbol{\rho}^T\right\} \boldsymbol{\Phi}^{-1/2} (\boldsymbol{\gamma}_t) \quad (82)$$

where

$$\boldsymbol{\gamma}_t = \begin{pmatrix} \bar{\boldsymbol{x}}_{t+\tau} \\ \bar{\boldsymbol{x}}_t \end{pmatrix}, \quad \boldsymbol{\rho} = \boldsymbol{\Phi}^{-1/2}\begin{pmatrix} \mathbf{V}\mathbf{W} \\ \mathbf{V} \end{pmatrix} \quad (83)$$

and

$$\boldsymbol{\Phi}^{-1} = \begin{pmatrix} (\mathbf{V}\boldsymbol{\Lambda}_\varepsilon\mathbf{V}^T + \boldsymbol{\Lambda}_x)^{-1} & \\ & (\boldsymbol{\Lambda}_x)^{-1} \end{pmatrix} \quad (84)$$

As shown in Eq. (82), the proof of **Theorem 3** is similar to that of **Theorem 2**. Hence, the detailed information is omitted for brevity. The monitoring policy for the model residual can be developed through

$$Q_{seq} = (\boldsymbol{\gamma}_t)^T \mathbf{H}(\boldsymbol{\gamma}_t) \leq \chi^2_{(1-\alpha,\kappa_s)} \quad (85)$$

where, $\mathbf{H} = \boldsymbol{\Phi}^{-1/2}\left\{\mathbf{I}_{2q} - \boldsymbol{\rho}(\boldsymbol{\rho}^T\boldsymbol{\rho})^{-1}\boldsymbol{\rho}^T\right\}\boldsymbol{\Phi}^{-1/2}$.

For this case, the model residual is developed by combining two adjacent samples. This is similar to the statistics developed in dynamic PCA (Ku, et al., 1995) and dynamic PLS (Chen, et al., 2002). Therefore, both static and dynamic information can be reflected in this monitoring index.

*4.3 Contribution Analysis for Monitoring Statistics*

Once a process anomaly is detected, the contribution of each variable to monitoring statistic should be calculated for fault diagnosis. Specifically, variables dominating the violation in monitoring statistic can be identified by evaluating their significance. The commonly used GDC and RBC are adopted as examples in this study, and their detailed information is analyzed based on the designed statistics.

All the monitoring statistics in GPMM model are quadratic forms, and thus they can be simplified by considering one general index as below

$$Index = \hbar^T \amalg \hbar \quad (86)$$

where, $\hbar_t = \left(\bar{\boldsymbol{y}}_t^T, \bar{\boldsymbol{x}}_t^T\right)^T$ and $\hbar_t = \left(\bar{\boldsymbol{x}}_{t+\tau}^T, \bar{\boldsymbol{x}}_t^T\right)^T$ for random data and sequential data, respectively. Besides, the covariance matrix $\amalg$ can be calculated accordingly.

The GDC method (Alcala, et al., 2011) decomposes a statistic as the summation of variable contributions, and it is defined as below

$$GDC_{spe} = \hbar^T \amalg^{1-\theta} \boldsymbol{\xi}_{spe} \boldsymbol{\xi}_{spe}^T \amalg^{\theta} \hbar \quad 0 \leq \theta \leq 1 \quad (87)$$

where, $\boldsymbol{\xi}_{spe}$ is a vector. Its element corresponding to the specific variable is 1, and other elements are 0.

Dividing each contribution by its expectation, and then the relative GDC (rGDC) can be further derived as below

$$rGDC_{spe} = GDC_{spe} / \boldsymbol{\xi}_{spe}^T \amalg^{\theta} \boldsymbol{\Psi}_h \amalg^{1-\theta} \boldsymbol{\xi}_{spe} \quad (88)$$

where, $\boldsymbol{\Psi}_h$ denotes the covariance matrix of $\hbar_t$.

It is noted that GDC reduces to partial decomposition contribution when $\theta = 0$ or $\theta = 1$, and reduces to complete decomposition contribution when $\theta = 1/2$.

The RBC method uses the amount of reconstruction of a fault detection index along a variable direction as the contribution of that variable. The detailed information can be found in Alcala et al. (2011), and the result is given as

$$RBC_{specific}^{Index} = \left(\boldsymbol{\xi}_{spe}^T \amalg \hbar\right)^2 / \left(\boldsymbol{\xi}_{spe}^T \amalg \boldsymbol{\xi}_{spe}\right) \quad (89)$$

Similarly, the relative RBC (rRBC) is given as

$$rRBC_{spe} = RBC_{spe} / \boldsymbol{\xi}_{spe}^T \amalg \boldsymbol{\Psi}_h \amalg \boldsymbol{\xi}_{spe} \quad (90)$$

## 5 Equivalent to Conventional Methods

*5.1 Equivalent to PCA and CCA Methods*

Raveendran et al. (2018) have proved that GPLLVM can be seamlessly reduced to monitoring strategies based on PPCA and PCCA. For monitoring model with random data, we will prove that the proposed GPMM can be reduced to GPLLVM under specific conditions.

**Lemma 3.** When $\mathbf{W} = \mathbf{I}_r$, $\boldsymbol{c}_x = 0$, and $\boldsymbol{c}_y = \mathbf{F}\boldsymbol{c}_t$, GPMM with random data reduces to GPLLVM.

**Proof.** Under the restrictions given in **Lemma 3**, the expectation and covariance of GPMM are reduced as below

$$\boldsymbol{\mu}_{s_t|\boldsymbol{x}_t,\boldsymbol{y}_t} = \boldsymbol{\Xi}_{s_t|\boldsymbol{x}_t,\boldsymbol{y}_t} \times \left[\mathbf{U}^T(\boldsymbol{\Lambda}_y)^{-1}\bar{\boldsymbol{y}}_t + \mathbf{V}^T(\boldsymbol{\Lambda}_x)^{-1}\bar{\boldsymbol{x}}_t\right] \quad (91)$$

$$\Xi_{s_t|x_t,y_t} = \left[\mathbf{U}^T(\mathbf{\Lambda}_y)^{-1}\mathbf{U} + \mathbf{V}^T(\mathbf{\Lambda}_x)^{-1}\mathbf{V} + \mathbf{I}_r\right]^{-1} \quad (92)$$

They are equivalent to the expectation and covariance of latent variable in Table A.1 of GPLLVM.

Combining with the results given PPCA (Tipping et al., 1999) and PCCA (Bach, et al., 2005), the equivalence between monitoring statistics from GPMM model and conventional monitoring models (i.e., PCA and CCA) can be proved. The detailed procedure is similar to that in Raveendran et al. (2018), and thus it is omitted in this study for brevity.

*5.2 Equivalent to SFA Method*

Shang et al. (2015) have proposed a PSFA method under the Markov chain architecture. It is easy to find that the GPMM in Eq. (20) can be reduced to PSFA under the restrictions $\tau = 1$ and $\mathbf{\Lambda}_x$ is a diagonal matrix. Since the parameters are estimated in a recursive manner, the equivalence between GPMM with sequential data and classical SFA based monitoring model is difficult to establish.

As an alternative, a GPMM model is developed to build connection with SFA based monitoring model using the first order difference of sequential data. As for this case, the GPMM is developed by setting $\mathbf{W} = \mathbf{\Lambda}_\varepsilon = \mathbf{I}_r$ and $\mathbf{c}_x = 0$, and its detailed information is given below

$$\begin{aligned} \mathbf{x}_{t+1} &= \mathbf{V}\mathbf{s}_{t+1} + \mathbf{e}_x, \quad \mathbf{e}_x \sim N(0, \mathbf{\Lambda}_x) \\ \mathbf{s}_{t+1} &= \mathbf{s}_t + \boldsymbol{\varepsilon}, \quad \boldsymbol{\varepsilon} \sim N(0, \mathbf{I}_r) \end{aligned} \quad (93)$$

In Eq. (93), the latent variable $s_t$ follows a random walk, and thus it is nonstationary. This is different from the time series depicted in Eq. (78), whose latent variable is a stationary sequence.

**Lemma 4.** Implementing whitening transformation to $x_t$, we can then obtain $\mathbf{z}_t = \mathbf{\Pi}^{-1/2}\mathbf{A}^T\mathbf{x}_t$ using Eq. (8). Define the first order difference of observation and latent variable as $\dot{\mathbf{z}}_t = \mathbf{z}_t - \mathbf{z}_{t-1}$ and $\dot{\mathbf{s}}_t = \mathbf{s}_t - \mathbf{s}_{t-1}$, respectively. The probabilistic interpretation for SFA can then be recast as below

$$\dot{\mathbf{z}}_t = \mathbf{V}\dot{\mathbf{s}}_t + \mathbf{e}_z, \quad \mathbf{e}_z \sim N(0, \mathbf{\Lambda}_z) \quad (94)$$

The parameters of Eq. (94) can be calculated using method which is depicted in Table I. For brevity, the final results are given as follows

$$\boldsymbol{\mu}_{s_t|z_t} = \Xi_{s_t|z_t}\mathbf{V}^T(\mathbf{\Lambda}_z)^{-1}\dot{\mathbf{z}}_t, \quad \Xi_{s_t|z_t} = \left[\mathbf{V}^T(\mathbf{\Lambda}_z)^{-1}\mathbf{V} + \mathbf{I}_r\right]^{-1} \quad (95)$$

Based on $\boldsymbol{\mu}_{s_t|z_t}$ and $\Xi_{s_t|z_t}$, two monitoring statistics which are similar to Eqs. (53) and (69) are designed as following

$$S_F^{Ran} = \left(\boldsymbol{\mu}_{s_t|z_t}\right)^T\left(\mathbf{I}_r - \Xi_{s_t|z_t}\right)^{-1}\left(\boldsymbol{\mu}_{s_t|z_t}\right) \quad (96)$$

$$S_S^{Ran} = \min_{\dot{s}_t}(\dot{\mathbf{z}}_t - \mathbf{V}\dot{\mathbf{s}}_t)^T(\mathbf{\Lambda}_z)^{-1}(\dot{\mathbf{z}}_t - \mathbf{V}\dot{\mathbf{s}}_t) \quad (97)$$

where, $\dot{\mathbf{s}}_t = \left[\mathbf{V}^T(\mathbf{\Lambda}_z)^{-1}\mathbf{V}\right]^{-1}\mathbf{V}^T(\mathbf{\Lambda}_z)^{-1}\dot{\mathbf{z}}_t$.

Both statistics $S_F^{Ran}$ and $S_S^{Ran}$ follow the chi square distribution, and this can be proved using the procedures similar to **Theorems 1 and 2**, respectively. Besides, the PGM model of Eq. (94) with the restriction $\mathbf{\Lambda}_z = \pi^2\mathbf{I}$ is identical to PPCA. Hence, the results of PPCA in Tipping & Bishop (1999a) are employed for the following analysis.

**Claim 1.** PGM model in Eq. (94) can be reduced to PPCA when setting $\mathbf{\Lambda}_z = \pi^2\mathbf{I}$. Hence, the detailed information presented in Section 3.2 of Tipping & Bishop (1999a) is employed as below

$$\mathbf{V}^{ML} = \mathbf{P}_r^T(\mathbf{\Omega}_r - \pi^2\mathbf{I}_r)^{1/2}\mathbf{R}, \quad \pi_{ML}^2 = \sum_{i=r+1}^{q}\lambda_i/(q-r) \quad (98)$$

where, the definition of $\mathbf{P}_r$ and $\mathbf{\Omega}_r$ are the first $r$ columns in matrices $\mathbf{P}$ and $\mathbf{\Omega}$, which can be found in Section 2.3. $\mathbf{R}$ is an arbitrary $r \times r$ orthogonal rotation matrix, and $\lambda_i$ $\{i = r+1, \cdots q\}$ indicate the minor eigenvalues of the covariance matrix $\mathbf{\Omega}$.

**Theorem 4.** When developing probabilistic SFA using Eq. (94) with restriction $\mathbf{\Lambda}_z = \pi^2\mathbf{I}_r$, under the maximum likelihood estimations in Eq. (98), the monitoring statistics $S_F^{Ran}$ (i.e., Eq. (96)) and $S_S^{Ran}$ (i.e., Eq. (97)) are equivalent to $S_F^2$ and $S_S^2$ (i.e., Eq. (12)), respectively.

**Proof.** Substituting Eq. (95) to Eq. (96), we have

$$S_F^{Ran} = \frac{1}{\pi^2}\dot{\mathbf{z}}_t^T\mathbf{V}\Xi_{\dot{s}_t|\dot{z}_t}\left(\mathbf{I}_r - \Xi_{\dot{s}_t|\dot{z}_t}\right)^{-1}\Xi_{\dot{s}_t|\dot{z}_t}\mathbf{V}^T\dot{\mathbf{z}}_t\frac{1}{\pi^2} \quad (99)$$

Using $\dot{\mathbf{s}}_t$ and Eq. (98), the statistic $S_S^{Ran}$ can then be depicted as below

$$S_S^{Ran} = \frac{1}{\pi^2}(\dot{\mathbf{z}}_t - \mathbf{V}\left[\mathbf{V}^T\mathbf{V}\right]^{-1}\mathbf{V}\dot{\mathbf{z}}_t)^T(\dot{\mathbf{z}}_t - \mathbf{V}\left[\mathbf{V}^T\mathbf{V}\right]^{-1}\mathbf{V}\dot{\mathbf{z}}_t) \quad (100)$$

Based on Eqs. (99) and (100), statistics $S_F^{Ran}$ and $S_S^{Ran}$ can be further reduced as below

$$S_F^{Ran} = \dot{\mathbf{x}}_t^T\mathbf{B}_F^T\Omega_F^{-1}\mathbf{B}_F\dot{\mathbf{x}}_t, \quad S_S^{Ran} = \frac{1}{\pi^2}\dot{\mathbf{x}}_t^T\mathbf{B}_S^T\mathbf{B}_S\dot{\mathbf{x}}_t \quad (101)$$

It is easy to find that statistic $S_F^{Ran}$ is identical to $S_F^2$ in Eq. (12). Besides, statistic $S_S^{Ran}$ should be scaled by $1/\pi^2$ when comparing with statistic $S_S^2$ in Eq. (12). Since the control limit of statistic $S_S^{Ran}$ should also be scaled with the same factor, the monitoring results will not be affected.

**6 Simulation Studies**

*6.1 A Numerical Example*

In this case study, the derived distributions of the statistics are verified, and the equivalence between GPMM method and classical methods is illustrated. A total of three simulated datasets are generated, including a random data and two sequential data. Specifically, the parameters in (Raveendran et al., 2018) have been adopted in this study to make a fair comparison. Besides, the elements in transition matrix $\mathbf{W}$ can be set to arbitrary values which satisfy $\lambda_i \in (0,1)$, and it is generated randomly in this case study. For all the monitoring statistics, two elements are retained in the latent variables for detecting the process anomalies.

The random data is simulated using Eq. (17) with parameters given in Eq. (102). The expectations of input $x_t$ and output $y_t$ are set as $\mathbf{c}_x = [0,0]^T$ and $\mathbf{c}_y = [0,0]^T$. For this case, $10^5$ samples are generated to illustrate the performance of statistics defined in Eqs. (53), (54), (69), (72), and (73).

The first sequential data is simulated using Eq. (78) with restriction $\tau = 1$. Specifically, the parameters $\mathbf{V}$, $\mathbf{\Lambda}_x$, $\mathbf{W}$ and

$\mathbf{\Lambda}_\varepsilon$ defined in Eq. (102) are adopted. The first latent variable satisfies $s_1 \sim N(0, \mathbf{I})$, and the expectations of $\mathbf{x}_t$ are set to $\mathbf{c}_x = [0, 0]^T$. A set of 100 process data is generated to train the GPMM model, and each of them contains 500 samples. Based on the obtained datasets, the performance of statistic defined in Eq. (85) is illustrated.

$$\mathbf{U} = \begin{bmatrix} 2.3 & -2.9 & 1.8 \\ 1.5 & 2.4 & -3.1 \end{bmatrix}^T, \quad \mathbf{V} = \begin{bmatrix} 1.2 & 3.2 & 1.3 \\ -2.3 & 1.7 & -2.4 \end{bmatrix}^T$$

$$\mathbf{\Lambda}_y = \begin{bmatrix} 0.8 & 0.2 & 0.3 \\ 0.2 & 0.5 & -0.4 \\ 0.3 & -0.4 & 0.9 \end{bmatrix}, \quad \mathbf{\Lambda}_x = \begin{bmatrix} 0.8 & 0.4 & 0.3 \\ 0.4 & 0.9 & -0.2 \\ 0.3 & -0.2 & 0.8 \end{bmatrix} \quad (102)$$

$$\mathbf{W} = \begin{bmatrix} 0.5400 & 0 \\ 0 & 0.6200 \end{bmatrix}, \quad \mathbf{\Lambda}_\varepsilon = \begin{bmatrix} 0.7084 & 0 \\ 0 & 0.6165 \end{bmatrix}$$

Different from the first sequential dataset, the second one is a nonstationary process. It is simulated using Eq. (93) with parameters $\mathbf{V}$ and $\mathbf{\Lambda}_x$ depicted in Eq. (102). Similar to the random data, a total of $10^5$ samples are generated in this process to illustrate the performance of statistics defined in Eqs. (96) and (97).

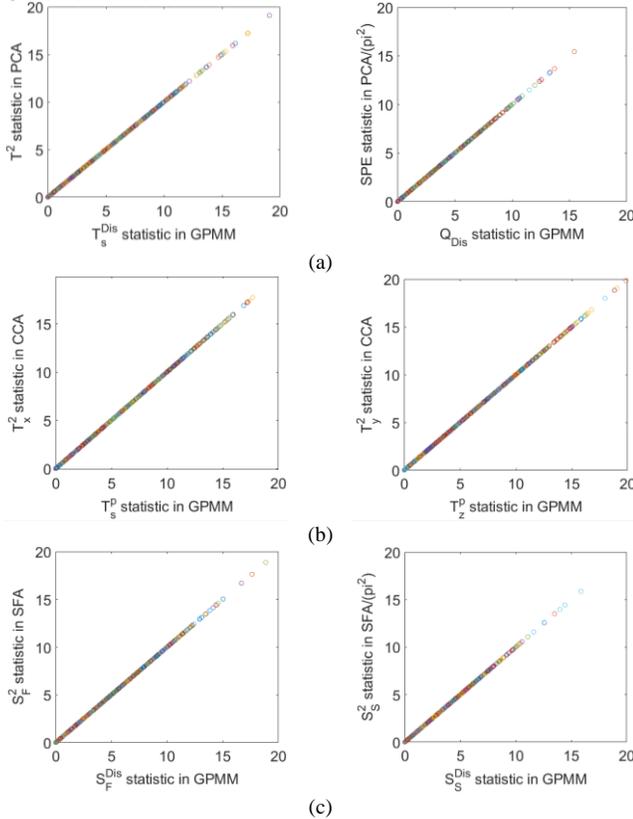

(a)

(b)

(c)

Fig. 3. Equivalence between monitoring statistics in GPMM and classical methods: (a) PCA, (b) CCA, and (c) SFA.

For monitoring statistic $Q_{Ran}$ given in Eq. (69), the eigenvalues of $\mathfrak{A}$ defined in Eq. (64) are $[1, 1, 1, 1, 0, 0]$, which is consistent with the conclusion that $\mathfrak{A}$ is an idempotent matrix. Since the rank of matrix $\mathfrak{A}$ is 4, the statistic $Q_{Ran}$ in Eq. (69) has a chi square distribution with 4 degrees of freedom. The simulations are repeated with 50 Monte-Carlo simulations, and the results of the developed statistics are summarized in Table II. As shown in this Table, the false alarm ratios of eight monitoring statistics are close to the desired values. That is, the developed statistics follow the chi square distribution with specific degree of freedom.

The equivalence between statistics in GPMM and classical methods is also illustrated using the generated data. Among them, the relationship between statistics in GPMM and these in PCA and CCA is verified using the random data. It is noted that the PPCA and PCA based monitoring models are developed using only the input data $\mathbf{x}_t$. Besides, the relationship between GPMM and SFA is validated using the second sequential data. The experimental results are depicted in Fig. 3, where the statistics of GPMM and PCA, CCA, and SFA are provided in Parts (a), (b), and (c), respectively. As shown in this figure, it is easy to find that the statistics derived from GPMM are identical to these of classical methods.

### 6.2 The Tennessee Eastman (TE) Process

In this subsection, the performance of the proposed method is further illustrated using TE process, which is a realistic simulation program of a chemical plant (Downs & Vogel, 1993). Specifically, 11 manipulated variables (XMV(1-11)) and 22 process measurements (XMEAS(1-22)) are employed to develop monitoring model since they are sampled every 3 minutes. TE process mainly contains four types of faults, including step, random variation, sticking, and unknown. For each type, a good and a bad monitoring results of GPMM method are summarized in Table III as representative. Since statistics $S_F^{Ran}$ and $S_S^{Ran}$ are used to reflect the control performance of the process system (Shang et al., 2015), the monitoring performance of them is omitted in this study. Besides, the monitoring results of PCA and DPCA in other literature (Shang et al., 2017; Yin et al., 2012) are adopted to provide a comparison.

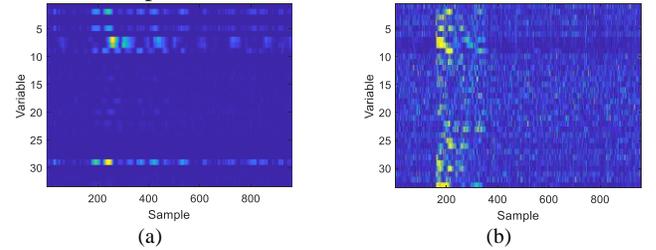

(a) (b)

Fig. 4. Contribution analysis of statistic $T_s^{Ran}$ for IDV(5): (a) GDC and (b) rRBC. (XMEAS(1-22) and XMV(1-11) are arranged as 1st-22nd and 23rd-33rd variables, respectively.)

The monitoring results summarized in Table III also illustrate the equivalence between GPMM and classical methods. As shown in this table, the false alarm ratios and detection rates of GPMM (especially for statistics $T_s^{Ran}$, $T_z^{Ran}$, $T_s^p$, and $T_z^p$) and classical methods (i.e., PCA and DPCA) are almost the same. Actually, the minor difference between them are caused by the model parameters selection and the control limits calculation. Since the Gaussian assumption may not be satisfied, kernel density estimation method is suggested as an alternative in real industrial applications.

In this study, the IDV(5) is taken as an example to illustrate the performance of the selected contribution analysis methods. For this case, a step disturbance is involved in condenser cooling water inlet temperature. As a consequence, many variables will be affected, including the final product flow (i.e., XMEAS(17)) and condenser cooling water flow (i.e., XMV(11)). The disturbance is introduced at 161st sample, and the control performance is completely recovered after 350th sample. The statistic $T_s^{Ran}$ is taken as an example, and its contribution plots of rGDC and rRBC are depicted in Fig. 4. As shown in this figure, it is easy to find that the starting and ending time of IDV(5) can be accurately reflected. In comparison with rGDC, XMEAS(9), (19), and XMV(9), (11) can be successfully recognized by rRBC. In this sense, rRBC has a better performance than rGDC for faulty variables identification of IDV(5).

## 7 Conclusions

In this study, a generalized probabilistic monitoring model is proposed for investigating the connection between classical multivariate methods and their corresponding probabilistic counterparts. The GPMM can be developed with both random and sequential data, and the solutions for these two cases are provided using EM algorithm. Based on the obtained model parameters, a series of statistics are designed for monitoring different aspects of the system, and their distributions are derived to calculate the corresponding control limits. Besides, GPMM can be reduced to many classical methods by setting restrictions to its model parameters. The equivalence between monitoring statistics of GPMM and these of PCA, CCA, and SFA is also analyzed and proved. The conclusions of this study are illustrated using a numerical example and the TE process. Experimental results verify that the established statistics follow the derived distributions, and illustrate the equivalence between statistics derived from GPMM and classical deterministic methods.

This study mainly focuses on developing GPMM model and investigating the equivalence between monitoring statistics from GPMM and classical methods. In our future work, the fault diagnosis methods which corresponding to the proposed method will be further studied. Besides, the proposed method will be also integrated with control systems to develop fault tolerant control system.

Table I. The estimations of model parameters in GPMM with random data

| E-Step: | $E_{\mathbf{X,Y,\Theta}^{old}}(\bar{x}_t s_t^T) = \bar{x}_t \times E_{\mathbf{X,Y,\Theta}^{old}}(s_t^T) = \bar{x}_t \times \mu_{s_t|x_t,y_t}^T$, $E_{\mathbf{X,Y,\Theta}^{old}}(s_t s_t^T) = \Xi_{s_t|x_t,y_t} + \mu_{s_t|x_t,y_t}\mu_{s_t|x_t,y_t}^T$ |
|---|---|
| | $E_{\mathbf{X,Y,\Theta}^{old}}(\bar{y}_t z_t^T) = \bar{y}_t \times E_{\mathbf{X,Y,\Theta}^{old}}(z_t^T) = \bar{y}_t \times \mu_{z_t|x_t,y_t}^T$, $E_{\mathbf{X,Y,\Theta}^{old}}(z_t z_t^T) = \Xi_{z_t|x_t,y_t} + \mu_{z_t|x_t,y_t}\mu_{z_t|x_t,y_t}^T$ |
| M-Step: | $\mathbf{V} = \left\{\sum_{t=1}^T E_{\mathbf{X,Y,\Theta}^{old}}(\bar{x}_t s_t^T)\right\}\left\{\sum_{t=1}^T E_{\mathbf{X,Y,\Theta}^{old}}(s_t s_t^T)\right\}^{-1}$, $\Lambda_x = \frac{1}{T}\sum_{t=1}^T \left\{(\bar{x}_t \bar{x}_t^T) - 2\left[E_{\mathbf{X,Y,\Theta}^{old}}(\bar{x}_t s_t^T)\right]\mathbf{V}^T + \mathbf{V}\left[E_{\mathbf{X,Y,\Theta}^{old}}(s_t s_t^T)\right]\mathbf{V}^T\right\}$ |
| | $\mathbf{U} = \left\{\sum_{t=1}^T E_{\mathbf{X,Y,\Theta}^{old}}(\bar{y}_t z_t^T)\right\}\left\{\sum_{t=1}^T E_{\mathbf{X,Y,\Theta}^{old}}(z_t z_t^T)\right\}^{-1}$, $\Lambda_y = \frac{1}{T}\sum_{t=1}^T \left\{E_{\mathbf{X,Y,\Theta}^{old}}(\bar{y}_t \bar{y}_t^T) - 2\left[E_{\mathbf{X,Y,\Theta}^{old}}(\bar{y}_t z_t^T)\right]\mathbf{U}^T + \mathbf{U}\left[E_{\mathbf{X,Y,\Theta}^{old}}(z_t z_t^T)\right]\mathbf{U}^T\right\}$ |

Table II. The false alarm ratios of the designed monitoring statistics

| No. | Monitoring statistics | Monitored aspect | $\alpha = 0.05$ | $\alpha = 0.01$ |
|---|---|---|---|---|
| 1 | $T_s^{Ran}$ in Eq. (53) | Variability in $s_t$ inferred from $x_t$ and $y_t$ with random data | $0.05 \pm 0.0008$ | $0.01 \pm 0.0009$ |
| 2 | $T_z^{Ran}$ in Eq. (54) | Variability in $z_t$ inferred from $x_t$ and $y_t$ with random data | $0.05 \pm 0.0006$ | $0.01 \pm 0.0006$ |
| 3 | $Q_{Ran}$ in Eq. (69) | Model residual with random data | $0.05 \pm 0.0012$ | $0.01 \pm 0.0003$ |
| 4 | $T_s^p$ in Eq. (72) | Variability in $s_t$ inferred from $x_t$ alone with random data | $0.05 \pm 0.0007$ | $0.01 \pm 0.0006$ |
| 5 | $T_z^p$ in Eq. (73) | Variability in $z_t$ inferred from $y_t$ alone with random data | $0.05 \pm 0.0009$ | $0.01 \pm 0.0008$ |
| 6 | $Q_{seq}$ in Eq. (85) | Model residual with sequential data | $0.05 \pm 0.0037$ | $0.01 \pm 0.0012$ |
| 7 | $S_F^{Ran}$ in Eq. (96) | Variability in $\dot{s}_t$ inferred from $\dot{y}_t$ with sequential data | $0.05 \pm 0.0019$ | $0.01 \pm 0.0008$ |
| 8 | $S_S^{Ran}$ in Eq. (97) | Model residual with sequential data | $0.05 \pm 0.0043$ | $0.01 \pm 0.0011$ |

Table III. False alarm ratios (i.e., normal process) and detection rates (i.e., faulty processes) for TE process

| | Type | $T_s^{Ran}$ | $T_z^{Ran}$ | $Q_{Ran}$ | $T_s^p$ | $T_z^p$ | $Q_{seq}$ | PCA (Yin et al., 2012) | DPCA (Shang et al., 2017) |
|---|---|---|---|---|---|---|---|---|---|
| Normal | -- | 6.26% | 4.90% | 4.80% | 4.80% | 5.11% | 6.15% | 6.13% | -- |
| IDV(1) | Step | 99.87% | 100.0% | 100.0% | 99.87% | 100.0% | 99.75% | 100.0% | 99.9% |
| IDV(5) | Step | 32.92% | 35.54% | 38.05% | 34.17% | 36.55% | 32.79% | 34.75% | 19.2% |
| IDV(8) | Random | 98.50% | 98.25% | 98.50% | 98.12% | 98.25% | 99.12% | 98.63% | 98.8% |
| IDV(10) | Random | 85.48% | 87.36% | 89.74% | 86.36% | 88.11% | 70.34% | 71.00% | 44.3% |
| IDV(14) | Sticking | 100.0% | 100.0% | 100.0% | 100.0% | 100.0% | 100.0% | 100.0% | 99.9% |
| IDV(15) | Sticking | 18.27% | 22.65% | 21.78% | 15.52% | 14.77% | 20.65% | 17.25% | -- |
| IDV(17) | Unknown | 92.87% | 97.25% | 97.87% | 94.62% | 96.75% | 94.37% | 96.88% | 98.2% |
| IDV(20) | Unknown | 70.34% | 70.34% | 87.36% | 69.84% | 72.22% | 68.21% | 71.50% | 58.0% |

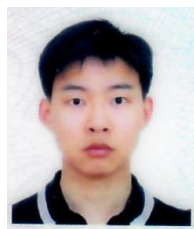
**Wanke Yu** received the B.Sc. degree in mathematics from Northeastern University, Shenyang, China, in 2013, the M.Sc. degree in pattern recognition from Beijing University of Aeronautics and Astronautics, Beijing, China, in 2016, and the Ph.D. degree in automatic control from Zhejiang University, Hangzhou, China, in 2020. He is currently a professor in the School of Automation, China University of Geosciences, Wuhan, China. His research interests include probabilistic graphic model, deep neural network, and nonconvex optimization and their applications to process control.

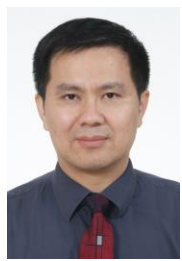
**Min Wu** (SM'08-F'19) received the B.S. and M.S. degrees in engineering from Central South University, Changsha, China, in 1983 and 1986, respectively, and his Ph.D. degree in engineering from the Tokyo Institute of Technology, Tokyo, Japan, in 1999.

He was a faculty member of the School of Information Science and Engineering at Central South University from 1986 to 2014, and was promoted to Professor in 1994. In 2014, he joined the China University of Geosciences, Wuhan, China, where he is currently a professor in the School of Automation. He was a visiting scholar with the Department of Electrical Engineering, Tohoku University, Sendai, Japan, from 1989 to 1990, and a visiting research scholar with the Department of Control and Systems Engineering, Tokyo Institute of Technology, from 1996 to 1999. He was a visiting professor at the School of Mechanical, Materials, Manufacturing Engineering and Management, University of Nottingham, Nottingham, UK, from 2001 to 2002. His current research interests include process control, robust control, and intelligent systems.

Dr. Wu is a Fellow of the IEEE and a Fellow of the Chinese Association of Automation. He received the IFAC Control Engineering Practice Prize Paper Award in 1999 (together with M. Nakano and J. She).

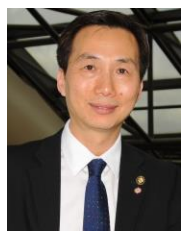
**Biao Huang** (M'97–F'18) received the B.Sc. and M.Sc. degrees in automatic control from the Beijing University of Aeronautics and Astronautics, Beijing, China, in 1983 and 1986, respectively, and the Ph.D. degree in process control from the University of Alberta, Edmonton, AB, Canada, in 1997.

He joined the University of Alberta, in 1997, as an Assistant Professor with the Department of Chemical and Materials Engineering, where he is currently a Professor and the NSERC Industrial Research Chair in Control of Oil Sands Processes. He has applied his expertise extensively in industrial practice. His current research interests include process control, system identification, control performance assessment, Bayesian methods, and state estimation.

Dr. Huang is a Fellow of the Canadian Academy of Engineering and the Chemical Institute of Canada. He was a recipient of Germany's Alexander von Humboldt Research Fellowship, the Canadian Chemical Engineer Society's Syncrude Canada Innovation and D. G. Fisher Awards, APEGA Summit Research Excellence Award, University of Alberta McCalla and Killam Professorship Awards, Petro-Canada Young Innovator Award, and a Best Paper Award from the Journal of Process Control.

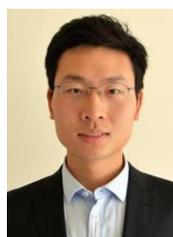
**Chengda Lu** received the B.S. degree in electronic information engineering from Wuhan University of Science and Technology, Wuhan, China, in 2012, the M.S. degree in information and communication engineering from China University of Geosciences, Wuhan, China, in 2015, and the Ph.D. degree in electrical and electronics engineering from Swinburne University of Technology, Melbourne, VIC, Australia, in 2019.

He joined China University of Geosciences, Wuhan, China, in 2019, where he is currently an Associate Professor with the School of Automation. His current research interests include robust control, time-delay systems, and neural networks.